\documentclass[letterpaper,twocolumn,10pt]{article}
\usepackage{usenix,epsfig,endnotes}
\usepackage{booktabs}
\usepackage{makecell}
\usepackage{tablefootnote}

\usepackage{cleveref}
\crefformat{section}{\S#2#1#3}
\crefformat{subsection}{\S#2#1#3}
\crefformat{subsubsection}{\S#2#1#3}

\usepackage[justification=centering]{caption}

\usepackage{longtable}

\makeatletter
\def\UrlAlphabet{%
      \do\a\do\b\do\c\do\d\do\e\do\f\do\g\do\h\do\i\do\j%
      \do\k\do\l\do\m\do\n\do\o\do\p\do\q\do\r\do\s\do\t%
      \do\u\do\v\do\w\do\x\do\y\do\z\do\A\do\B\do\C\do\D%
      \do\E\do\F\do\G\do\H\do\I\do\J\do\K\do\L\do\M\do\N%
      \do\O\do\P\do\Q\do\R\do\S\do\T\do\U\do\V\do\W\do\X%
      \do\Y\do\Z}
\def\UrlDigits{\do\1\do\2\do\3\do\4\do\5\do\6\do\7\do\8\do\9\do\0}
\g@addto@macro{\UrlBreaks}{\UrlOrds}
\g@addto@macro{\UrlBreaks}{\UrlAlphabet}
\g@addto@macro{\UrlBreaks}{\UrlDigits}
\makeatother

\begin{document}

%don't want date printed
\date{}

%make title bold and 14 pt font (Latex default is non-bold, 16 pt)
\title{\Large \bf ``I Cannot Write This Because It Violates Our Content Policy'':\\ Understanding Content Moderation Policies and User Experiences in \\Generative AI Products }

\author{
{\rm Lan Gao}, {\rm Oscar Chen}, {\rm Rachel Lee}, {\rm Nick Feamster}, {\rm Chenhao Tan}, {\rm Marshini Chetty}\\
University of Chicago\\
%\{langao, oscarc, rlee20, feamster, chenhao, marshini\}@uchicago.edu
}

\newif\ifcomments

\commentstrue

\ifcomments
  \newcommand{\chenhao}[1]{\textcolor{magenta}{\textsc{#1 ---CT}}}
\else
  \newcommand{\chenhao}[1]{}
\fi

\newcommand{\lan}[1]{\textcolor{blue}{\textsc{#1 ---LG}}}

\newcommand{\mc}[1]{\textcolor{red}{\textsc{#1 ---MC}}}

\maketitle

% Use the following at camera-ready time to suppress page numbers.
% Comment it out when you first submit the paper for review.
\thispagestyle{empty}

\subsection*{Abstract}
While recent research has focused on developing safeguards for generative AI (GAI) model-level content safety, little is known about how content moderation to prevent malicious content performs for end-users in real-world GAI products. To bridge this gap, we investigated content moderation policies and their enforcement in GAI online tools --- consumer-facing web-based GAI applications. We first analyzed content moderation policies of 14 GAI online tools. While these policies are comprehensive in outlining moderation practices, they usually lack details on practical implementations and are not specific about how users can aid in moderation or appeal moderation decisions. Next, we examined user-experienced content moderation successes and failures through Reddit discussions on GAI online tools. We found that although moderation systems succeeded in blocking malicious generations pervasively, users frequently experienced frustration in failures of both moderation systems and user support after moderation. Based on these findings, we suggest improvements for content moderation policy and user experiences in real-world GAI products.

\section{Introduction}

%\chenhao{is this our own definition? What do we mean by advanced?} \lan{Yes, and I removed `advanced' to prevent controversies in definition} \mc{we should probably make it clear that this is how we are defining it and maybe also explain how it is the same or differs from other definitions and add citations} \lan{This concept is what I summarized and unified from literature and government definition -- there is no consensus on the definition of AIGC. I add all my references in footnote} 
The development of large generative AI (GAI) models, such as large language models (LLMs) and diffusion models, has promoted the production of \textit{AI-Generated Content (AIGC)} --- synthetic content, in the form of text, image, audio, video, that are generated by AI models given human instructions.\footnote{There is no consensus on the definition of AIGC, and therefore the definition we used in this paper is summarized from existing literature (e.g., \cite{cao2023comprehensive, wu2023ai}) and related definitions in law \cite{uscode_title15_section9401}.} Recently, this has opened up AIGC as a new form for content creation as opposed to human-created content~\cite{cao2023comprehensive, wu2023ai, wei2024understanding, hua2024generative}. 

Simultaneously, the potential harms associated with GAI's generation capabilities, such as producing disturbing, misleading, and privacy/copyright-infringing AIGC~\cite{chen2023pathway, zhou2023synthetic,samuelson2023generative}, have raised attention. Malicious content generated by GAI tools could not only directly compromise the safety of end-users who interact with the system, but also pose security and privacy risks to the public due to its automated and fast generation. In response, AI and Security practitioners have been working on GAI safeguards, such as safety alignments and content filters, to prevent problematic content generation (e.g. \cite{markov2023holistic, ji2024beavertails, schramowski2023safe, inan2023llama}). 
%Their common approaches include model fine-tuning, input and output filtering, and detective and defensive mechanisms of malicious prompting (i.e., jailbreaking)~\cite{}. 
When deploying GAI models into real-world products, service providers also enforce \textit{content moderation} on users' content generation process. Content moderation is a common strategy employed in online communities to reduce problematic user-generated content by detecting and restricting such content and users who publish it~\cite{kiesler2012regulating, roberts2019behind}, and has now been used for GAI products. For example, ChatGPT denied over 250,000 requests for generating images of US political campaigns before the 2024 US election day, to prohibit potential misinformation~\cite{cnbc_chatgpt_image_generations}.

%On the other hand, online GenAI services, including general GenAI services like ChatGPT and AIGC-specified services like Midjourney, have established moderation mechanisms outside the model in addition to internal mechanisms. These services moderate user input (prompts) and AI output through automatic and human moderators

%Content moderation is undoubtedly vital for GAI products to limit malicious AIGC production and ensure user safety.

Nevertheless, most recent works focus on improving safeguards to GAI content generation safety at the \textit{model level}. Little is known about how and how well content moderation practices are enforced at the \textit{product level} (i.e., real-world GAI products, like ChatGPT). Since GAI is an emerging technology, there has been a lack of legal guidance on how GAI products should engage in content moderation until recently~\cite{hacker2023regulating}, with service providers instead relying on themselves for making policies~\cite{klyman2024acceptable}.
Therefore, it is crucial to understand what GAI tool content moderation policies are established by service providers. As seen in online communities~\cite{schaffner2024community, singhal2023sok}, content moderation policies reveal how service providers enforce moderation, disclose their practices to the public, and guide user behavior. Meanwhile, given the strong correlation of content moderation with user behavior and experiences~\cite{ma2023users, myers2018censored}, user perspectives on content moderation provide valuable insights into how current policy enforcement succeeds or fails in practice. Both processes together could inform the future direction of content moderation policies and practices for GAI products.

\begin{figure*}[htbp]
    \centering
    \includegraphics[width=0.97\linewidth]{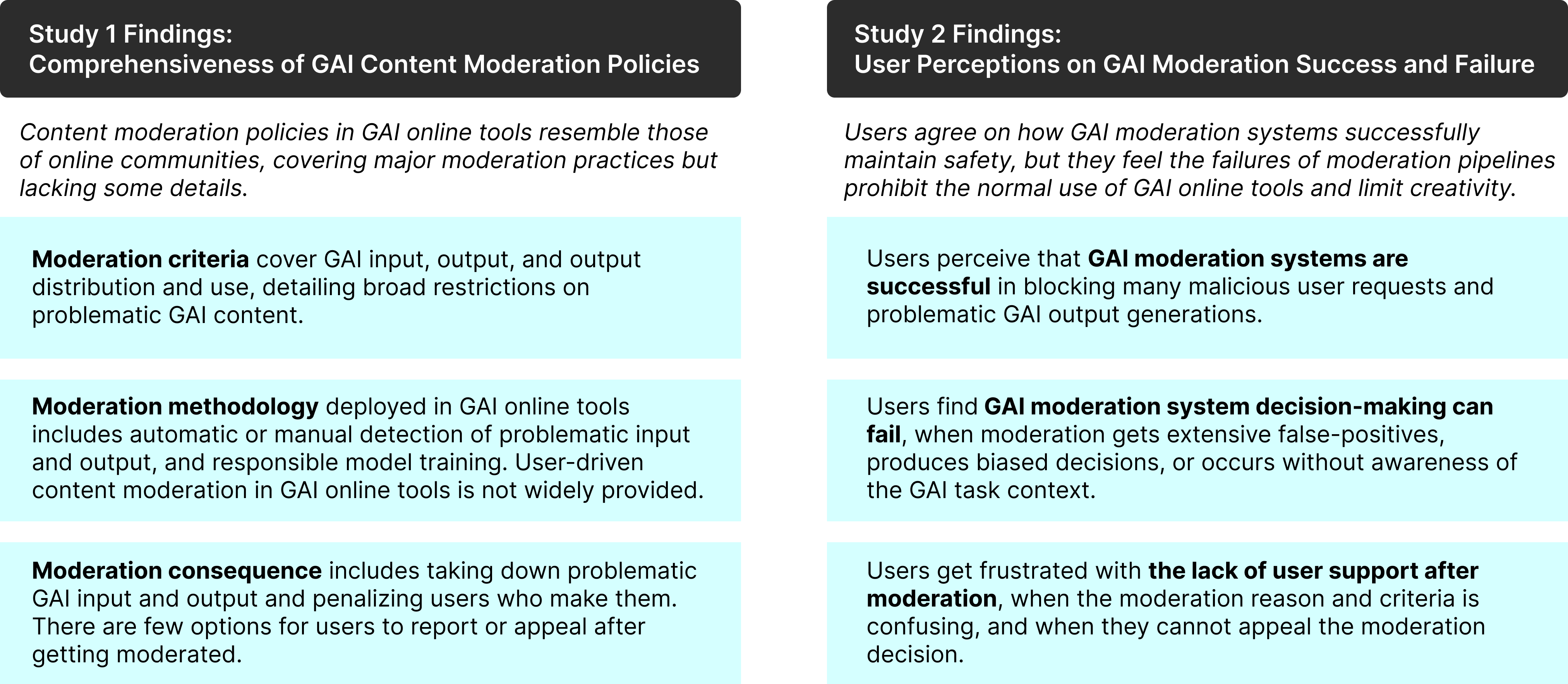}
    \caption{Findings summary of Study 1: Policy Analysis and Study 2: Reddit Study.}
    \label{fig:enter-label}
\end{figure*}

We study content moderation policies and their enforcement in consumer-facing GAI products offered through web-based applications (e.g., ChatGPT and its
playground, referred to as \textit{GAI online tools} from here on) that enable
text and image generation. GAI online tools are widely adopted by end-users
regardless of their technical expertise, as compared to locally deployed GAI
models and APIs. We focus only on text and image generation tools since they
make up many of the GAI online tool markets, despite the growth of video and
audio generation tools recently~\cite{statista_generative_ai_usage,
writerbuddy_ai_industry_analysis}. Through this work, we study the following
research questions:

\begin{itemize}
    \item RQ1 \textit{(Service Provider Policymaking)}: What themes do content
        moderation policies in GAI online tools cover, and how comprehensive
        are they? %\mc{this should change to be how comprehensive are the policies right now }
    \item RQ2 \textit{(User Experience on Policy Enforcement)}: From user
        perspectives, in what areas does content moderation policy enforcement
        for GAI online tools succeed, and in what areas do they fail? 
    %\chenhao{if we only look at the reddit sample, people are more likely to complain, so it is not representative.} \lan{I'm thinking of asking `According to user experience in AIGC creation, how content moderation policy enforced in GAI online tools does or does not work.' To that end, findings 2 will only include `the content moderation system did work on blocking malicious AIGC creation' in Section 7.3 (although very short) and lots of user criticisms in Section 7.2. Indeed users always feel bad when getting moderated. So we can just focus on how users think the regulation and moderation pipeline provided by service providers are flawed.} \mc{I think we need to address the comment Chenhao has more directly and say that we examine Reddit perspectives on the tools so that it is more clear this is not representative } \lan{Leave RQ2 here and would appreciate it if Chenhao and Nick have any idea on RQ2 and the framing of study 2 -- see the discussions above}
\end{itemize}

To answer RQ1, we studied the content moderation policies of 14 GAI online
tools that were capable of text and image generation (Study 1: Policy
Analysis). We found that content moderation policies in GAI online tools are
similar to those in online communities~\cite{schaffner2024community}, with
scattered locations in different pages outlining three components of content
moderation practices, including moderation criteria on forbidden content,
methodology to detect problematic content, and the consequences of problematic
content and users.  Compared to how online communities rule on user-generated
content, policies in GAI online tools focus on governing both user input to
the tool and content generated by the tool, with some differences in
moderation methodology and consequences for policy violations (See Table
\ref{tab:policycomparision}). We found that GAI online tools are comprehensive in
covering major content moderation practices, but these policies lack details
on topics such as how users can report problematic content generation and appeal moderation decisions.

To address RQ2, we studied user experiences with content
moderation while using GAI online tools to generate new, creative, or crafted
AIGC (e.g., requesting ChatGPT to write a poem about cats or using its DALL-E
to draw a cat, referred to as \textit{AIGC creative tasks} from here on).
Tasks performed through GAI are diverse: LLMs can generate text for
dialogue, question-answering, searching, and so on. These tasks result in
varied presentations of generated content (e.g., natural speech in dialogues
versus structured writing in creative writings), which may affect moderation
performance~\cite{mahomed2024auditing} and influence user mental model on
content moderation. Therefore, we studied the AIGC creative tasks in
particular, which is not only a representative use case of
GAI~\cite{statista_generative_ai_us}, but also heavily depends on GAI's
generation capability compared to other usages like searching.

To do so, we analyzed public discussions on content moderation in AIGC
creative tasks when using GAI online tools, by looking into user posts and
corresponding comments on Reddit covering six different GAI online tools
(Study 2: Reddit Study). We found that while user-provided examples
highlighted the widespread success of moderation systems in blocking malicious
AIGC creation, users also discussed numerous instances of moderation pipeline
failures. These failures spanned both moderation systems' shortcomings in
making justified moderation decisions, and lack of support for users to
understand and appeal moderation decisions. As users posted, failures in the
moderation pipeline had severely hampered the capability and usability of GAI
online tools, limiting their creativity in AIGC creative tasks as well. See Figure \ref{fig:enter-label} for key findings of the two studies.

Our work provides in-depth insights into the content moderation pipeline in real-world GAI products, a key feature for controlling malicious content generation and user behavior to safeguard security and privacy. Specifically, our work makes the following contributions in AI security, privacy, and safety domains: (1) we provide the first study of GAI products' content moderation policies and existing gaps,
(2) we compile two datasets of policies and public discussions on GAI products' content moderation for further studies,\footnote{The datasets are publicly available at \url{https://doi.org/10.6084/m9.figshare.29257187}} and
(3) we suggest how to improve policies and user experiences with content moderation when using these products.  

 \vspace{-1.3em}
\section{Related Works}
In this section, we review three sets of prior research that are relevant to our study, highlighting how our work extends or contributes to each body of research.
%: safeguards in GAI for ensuring content safety, content moderation policies and GAI, and user experience with content moderation. We discuss how our work extends and contributes to each body of research.

 \vspace{-1.5em}
\paragraph{Safeguards in GAI for Ensuring Content Safety.} AI and security researchers use two automated methods to ensure the safety of GAI model output: internal model fine-tuning and external content guardrails. Prior works have shown the success of widely adopted reinforcement learning methods for LLMs in providing targeted feedback on output toxicity through fine-grained rewards \cite{wu2023finegrainedhumanfeedbackgives, schulman2017proximalpolicyoptimizationalgorithms, ji2024beavertails}. Recently, AI alignment, also known as Reinforcement Learning from Human Feedback
(RLHF), has made further advancements in ensuring desirable model output. This method evaluates the harmfulness and helpfulness of model output with the guidance of human feedback~\cite{ji2023ai}. 
Following this principle, researchers have enforced safety alignment in LLMs to provide harmless responses and reject problematic model input \cite{ji2024beavertails, ouyang2022traininglanguagemodelsfollow, bai2022traininghelpfulharmlessassistant, dai2023saferlhfsafereinforcement}. 
Researchers have also focused on adjusting multi-modal generation models for generating safe output~\cite{liu2024safetydpo, dai2024safesora, wang2024moderator, schramowski2023safe}. %For instance, Schramowski et al. developed Safe Latent Diffusion for text-to-image models that modify the latent space during generation to suppress inappropriate output images \cite{schramowski2023safe}.

%Due to commonly known issues where LLMs struggle with poor calibration when handling novel inputs, LLMs can generate inappropriate outputs based on bias or no outputs at all \cite{cercas-curry-rieser-2018-metoo, lee-etal-2019-exploring}. 
%Additionally, prior reinforcement learning models have been trained on comparisons of trajectory segments to coordinate model behavior with human intent, reducing the need for predefined reward functions and extensive human oversight and thus enabling reinforcement learning systems to learn complex tasks such as Atari games and robotics behaviors with limited human feedback \cite{christiano2023deepreinforcementlearninghuman}.

Aside from internal model fine-tuning, content guardrails are also deployed outside the GAI model to monitor both model input and output. 
Similar to traditional algorithmic moderation in online communities, some guardrails in GAI apply classifiers trained with categorically labeled content on different harms~\cite{markov2023holistic, rando2022red}. While other state-of-the-art guardrails used classifier-free guidance --- for example, Llama Guard leverages instruction-tuned LLMs and safety taxonomies for problematic content detection~\cite{inan2023llama, wei2022finetunedlanguagemodelszeroshot, touvron2023llama2openfoundation}. 

%Existing research emphasizes the challenges of standardizing moderation taxonomies across applications and datasets. For example, prior works faced issues such as a lack of clear mappings for categories like self-harm, which highlights the concern of subjectivity in moderation policy 
%Content moderation in GAI has become a critical area of study due to its ability to mitigate harmful outputs. Previous work has explored both the technical and policy implications of content moderation, particularly in text generation and text-to-image (T2I) systems, with a focus on the challenges of balancing safety, transparency, and creativity.
%For instance, an audit of OpenAI’s moderation endpoint for GPT models observed high rates of flagged content for both real and generated outputs \cite{mahomed2024auditing, riccio2024exploring}. 

Through improving both internal and external safeguards, previous research has made significant strides in mitigating undesirable outputs within model architectures. However, there is scarce work investigating how content moderation is enforced with end-users in real-world GAI products, except for several works that audited the content moderation endpoint of GAI products~\cite{mahomed2024auditing, riccio2024exploring}
To bridge this gap, our work unpacks content moderation practices by examining GAI tool policies and public discussions where users talk about content moderation successes or failures in GAI products. 

%This includes analyzing policy enforcement decisions, user perceptions of fairness, and the interaction between moderation criteria and product-facing consumers. For example, we identify instances where moderation systems fail to account for the context of creative tasks, leading to inconsistent or overly restrictive enforcement that hampers usability.

%Our work also addresses gaps in transparency by investigating how moderation decisions are communicated—or not communicated—to users. Through our qualitative analysis, we are able to examine the effectiveness of user appeals, clarity in moderation criteria, and the challenges posed by ambiguous or incomplete policy language. In doing so, we are able to bridge the technical and policy aspects of AIGC moderation which could potentially inform both future technical safeguards and GAI policy development.
\vspace{-1em}
\paragraph{Content Moderation Policies in Online Communities and GAI.} Much research to date focuses on measuring and analyzing content moderation policies in online communities with user-generated content --- studying what types of content are forbidden, and how these rules vary around different platforms~\cite{arora2023detecting, buckley2022censorship, 10.1145/3406865.3418312, singhal2023sok, 10.1145/2818048.2819931, fiesler2018reddit}. Meanwhile, some researchers also analyzed the comprehensiveness of content moderation policies --- how and how much online communities disclose content moderation rules in their policies~\cite{singhal2023sok,schaffner2024community}. Schaffner et al., for example, conducted a mixed-method analysis of the comprehensiveness of content moderation policies across 43 platforms hosting user-generated content. Their findings show that while most policies articulate moderation goals, criteria, and practices, there is considerable variation in structure, composition, and legal grounding across platforms. They also found critical shortcomings of these policies, such as the lack of a clear definition of what to moderate, and the absence of user appeals for most moderation cases~\cite{schaffner2024community}.

Content moderation policies are informed by laws (e.g., The First Amendment and Section 230 in the United States context) while dominated by online communities themselves. Some works have focused on understanding how content moderation policies balance legal requirements, platform motivations, social values, and user experiences \cite{gillespie2020expanding, gillespie2017governance, suzor2019lawless, klonick2017new, gillespie2018custodians, goldman2021content, langvardt2017regulating}. 

More recent studies have recognized that regulating content generation in GAI products is an emerging issue worthy of attention~\cite{hacker2023regulating, schmitt2024implications,appel2024generative}. 
%Hacker et al., for example, highlighted the challenge of regulating GAI products content moderation, as legal frameworks for content moderation in online communities, such as the Digital Service Act (DSA) in the EU, were not applicable~\cite{hacker2023regulating}.
Some research works have examined what input and output in GAI usage are forbidden by policies and guidelines~\cite{klyman2024acceptable, riccio2024exploring}. Our work, furthermore, investigates the comprehensiveness of content moderation policies -- if and how these policies outline content moderation rules and practices.

\vspace{-1.3em}
\paragraph{User Experience with Content Moderation in Online Communities.} Prior research has investigated user experiences and reactions to content moderation in online communities~\cite{ma2023users}. Some works examined user understanding of content moderation~\cite{myers2018censored, 10.1145/2675133.2675234, 10.1145/3359294}, where researchers found users mainly rely on their own interpretation of how the moderation systems work. Prior studies also investigated user behaviors after moderation~\cite{10.1145/3359252}, and how users circumvent or interact with moderation systems~\cite{ma2021advertiser, moran2022folk}. Since policy statements may not align with actual practices~\cite{10.1145/3512965}, prior studies often rely on understanding user experiences to assess whether content moderation policy enforcement is effective or failing~\cite{singhal2023sok}. By measuring user behaviors in online communities on a large scale, researchers found content moderation successful in mitigating disturbing content (e.g., \cite{10.1145/3134666}).  Simultaneously, researchers also found common failures such as biases and inconsistencies of algorithmic moderation~\cite{lyons2022s, vaccaro2020end, haimson2021disproportionate,10.1145/3476059}; lack of transparency in moderation decisions~\cite{myers2018censored, juneja2020through}; and ineffectiveness of user appeals~\cite{myers2018censored}. Researchers also conducted user studies on user preferences and expectations of content moderation to inform the future content moderation policies and practices~\cite{10.1145/2818048.2819931, vaccaro2020end, 10.1145/3476059}.

Inspired by prior works investigating users in online community content moderation, we looked at how users experience content moderation in GAI products to understand the successes and failures of policy enforcement. Unlike previous studies, which typically focus on a single platform, our research also spans multiple GAI products, providing a broader and more generalized perspective.
% \paragraph{1. How users understand and react to content moderation}
% e.g. \cite{vaccaro2020end}(user appeal); \cite{myers2018censored}

% \paragraph{2. user-perceived good and bad thing of policy enforcement in online communities: effectiveness, bias, etc. in content moderation} 

% \paragraph{Research gap we bridge} Regardless of how content moderation correlated to user experience, which has been widely studied in the context of the online community, we do not know the user experience of content moderation in GAI. Taking the case of AIGC creation-oriented tasks, our work provides initial insight on user perception of content moderation in GAI online tools.

% \paragraph{Literature search and writing sample can be referred to:} https://dl.acm.org/doi/pdf/10.1145/3613904.3642333 -- P4: User understanding of content moderation

% Survey paper where you can supplement literature if needed: https://dl.acm.org/doi/pdf/10.1145/3610069
% https://ieeexplore.ieee.org/abstract/document/10190527 - section 3.2

\section{Study 1: Policy Analysis Methodology}
%Content moderation \mc{in the intro we may need to define what we mean by content moderation and include citations} is usually a black-boxed action in consumer-facing services. Content moderation policies, meanwhile, manifest how service providers define and regulate content moderation, which also serves as a medium to disclose content moderation practices to the public and end-users. \lan{TODO: maybe rewrote this part on the justification of the contribution of policy analysis}

To answer our first research question, we analyzed content moderation policies of 14 representative GAI online tools capable of text and image generation, inspired by and partly following the approach of Schaffner et al., who analyzed the comprehensiveness of content moderation policies in online communities~\cite{schaffner2024community}. For each tool, we manually located its content moderation policies, collected web pages where policies were situated, and qualitatively analyzed those policies. 

Regulated by regional laws, content moderation policies could be different when providing service in different regions. For example, content moderation of online communities is regulated by the First Amendment in the United States (US) but by the Digital Services Act (DSA) in the European Union (EU). Divergences such as DSA restricting protected speech defined by the First Amendment could lead to distinct policy enforcement between platforms in the US and EU~\cite{tourkochoriti2023digital, ahn2023splintering}. Acknowledging that regional policy alterations are out of our study scope, we only considered tools with US-based headquarters and accessed all policies through US IP addresses. 

%\mc{do we have an example or citation for this? what kinds of regulation and what kinds of tweaks?}

%We present details of our approach in the following section.
\subsection{Tool List Creation}
%When researching the GAI online tool market, we noticed that there had been no official reports on representative products. Some reports ranked the popularity of GAI products by measuring internet traffic, which is also a widely adopted method of measuring website popularity in research (e.g., Tranco \cite{pochat2018tranco}). However, these reports may not be reliable as sole references. Since the market of GAI products is changing fast, recent rising products are not covered in previous reports. Therefore, we self-curated a tool list of representative GAI online tools by referring to multiple resources (see below) and through discussions within the research team.

%\mc{add section reference so folks know we will add citations to the sources}
\label{sec:toolist}
We referred to multiple resources to create a representative tool list. First, we relied on a report \cite{writerbuddy_ai_industry_analysis} on popular GAI products, 
which has been used in other academic research papers (e.g., \cite{garcia2024generative}) and featured by Forbes News \cite{forbes_chatgpt_ai_tool_sector}.
%, indicating its good recognition among researchers and media workers. 
We picked out GAI online tools in the top 15 for text or image generation and with US-based headquarters, resulting in seven qualified tools. 
We excluded Character.AI,\footnote{https://character.ai/} a tool designed for role-playing, which may lead to a different policy focus compared to other multi-tasked or creativity/productivity-focused tools. This process resulted in seven GAI online tools.

We then supplemented our list from a user-curated GAI tool list (i.e., Awesome List \cite{github_awesome_generative_ai}),
survey and review papers on GAI and AIGC \cite{cao2023comprehensive, chen2023pathway}, another GAI tool popularity report~\cite{flexos_generative_ai_top150}, and AI products produced by big technology companies. We initially selected five tools through this process, which constructed our initial tool list along with the seven tools above. As the study progressed, we incorporated two additional tools based on ongoing tracking of supplemental resources.

%\mc{may want to think about if this compromises anonymity} \lan{This can be a position statement so should be fine?}
Our final list contains 14 GAI online tools, with two for text generation: Claude and You.com; five for image generation: CivitAI, Craiyon, DreamStudio, Firefly, and Midjourney; and seven capable of multi-modal generation of both text and image: ChatGPT, Copilot, Gemini, Meta AI, NovelAI, Perplexity AI, and xAI.

% \begin{table}[ht]
%     \centering
%     \small
%     \renewcommand\arraystretch{1.2}
%     \resizebox{1\columnwidth}{!}{
%     \begin{tabular}{|lp{0.35\textwidth}|}
%         \hline
%         \textbf{Generation Type} & \textbf{Tool}\\
%         \hline
%         Text & Claude, You.com\\
%         \hline
%         Image & CivitAI, Craiyon, DreamStudio, Firefly, Midjourney, \\
%         \hline
%         Multi-modal & ChatGPT, Copilot, Gemini, Meta AI, NovelAI, Perplexity AI, xAI\\
%         \hline
%     \end{tabular}}
%     \captionsetup{justification=centering}
%     \caption{Tool list for policy analysis}
%     \vspace{-0.3em}
%     \label{tab:toollist}
% \end{table}

\subsection{Content Moderation Policies Collection}
\label{subsec:policycollection}
Using the procedure outlined below, we collected content moderation policies in September 2024 for tools except Meta AI and xAI---two tools added in the middle of the study----and in December 2024 for Meta AI and xAI.

\textbf{Defining Policy Scope.} %Before starting to localize content moderation policies, we certified the policy scope that we wanted to analyze. We notice that many service providers also offer GAI services through APIs and open-sourced models to developers and enterprises, which is not in line with our research scope and questions. Therefore, 
We decided to only consider content moderation policies under our research scope --- governing activities where users directly interact with GAI through web-based applications. 
%\mc{define online playground and add citation} 
For example, OpenAI offers GPT API \footnote{https://openai.com/api/} to developers and allows people to tailor personalized GPT \footnote{https://openai.com/index/introducing-gpts/} and provide it to other users. Under this context, there are policies restricting the practices of developers and their customers, which are out of our scope.

%When studying policies, keyword-based searching is the most common approach to locate and collect related policies by sentences (e.g., \cite{}). We, however, used a more open-ended, formative way by manually reviewing all policy and support pages and recording relative ones. We adopted this method for two reasons. First, unlike privacy policies that are usually situated in a few 'Privacy Policy' pages, content moderation policies are usually not well-defined and segregated into a bunch of policy pages~\cite{}, leading to difficulties in locating policies in a few pages through keywords. Second, unlike content moderation in social media, where most platforms follow existing legal provisions to define their content policy, content moderation in GAI online tools lacks legal standardization to guide common practice, making it hard to locate policies comprehensively through existing keywords. Therefore, we believe our approach could perform better in exploring where, what, and how content moderation policies are in GAI online tools.

%\mc{add schaffner et al. to make sure the data gathering process is also seen as tried and tested}
\textbf{Locating Pages with Content Moderation Policies.} We followed a process used by Schaffner et al. \cite{schaffner2024community} %in their initial exploration of content moderation policy locations in online communities, 
to identify pages in GAI online tools that contained content moderation policies. For each tool, we manually examined all policy and regulation-related pages and collected two types of pages. First, we selected pages that included policies regulating user interactions with GAI. %However, we excluded policies that were unrelated to the process of users interacting with GAI (e.g., \textit{Subscription Policy}). 
Second, for those tools provided by companies developing multiple products (e.g., Google which developed Gemini), we also included company's general \textit{Terms of Service} (ToS) that applied to all products.

Next, we visited tool support pages, including the \textit{Help Center} and \textit{Frequently Asked Questions} (FAQ), if they existed. To identify information related to content moderation policies, we referred to Schaffner et al.'s four common elements of content moderation policies for user-generated content: what is moderated, why content is moderated, how the process of content moderation manifests, and who takes the responsibility for content moderation~\cite{schaffner2024community}. We then manually checked all support pages and their child pages, collecting those with information that fell into the above four themes. Additionally, we collected pages with information on system errors, which could also relate to content moderation enforcement.

\textbf{Recording Pages.} We saved all static pages to PDF through the MacOS Safari browser, where the web pages retain their original visual appearance and content, with all text editable in the converted PDF file. For those non-static pages in which the text encoding could not be captured through PDF conversion, we took screenshots of the whole pages. We used the Optical Character Recognition (OCR) service on the screenshots to convert all characters into editable text and save them in PDF format. We eventually recorded 52 PDF files of 51 pages, forming the policy dataset we analyzed. 

\subsection{Policy Analysis}
\label{susec: analysis1}
To get insights into how content moderation was defined and disclosed in GAI online tools' policies, we conducted a qualitative analysis of the policy dataset. Our analysis was performed deductively with multiple rounds using the analysis tool MAXQDA.\footnote{https://www.maxqda.com/} To start, the first author developed an initial codebook for the analysis and discussed it with the research team. The design of our initial codebook was based on four key components of how policies describe content moderation, as identified by Schaffner et al. \cite{schaffner2024community}: what content is moderated, why content is moderated, how content moderation is enforced, and who is responsible for content moderation. Using the initial codebook, the first author read all documents and deductively coded all segments that fit into existing themes. Simultaneously, excerpts that related to content moderation but did not fit the initial codebook were tagged with open codes (i.e., adding and removing themes) to build the final codebook. 

A second round of coding was then performed by two additional researchers using the final codebook. All documents were equally divided into two sets and assigned to these coders, ensuring that every document was coded by at least two coders at the end. Three coders met regularly to discuss the analysis process, compare each other's codes, and solve coding disagreements. Since our data informed the iterative analysis and we took care to minimize subjectivity and disagreement in the process, we did not calculate the inter-rater reliability (IRR)~\cite{10.1145/3359174, armstrong1997place}.

\textbf{Locations of Content Moderation Policies.} Within 51 pages recorded, 44 of them were coded during policy analysis, indicating that these pages contain content moderation policies. Similar to where content moderation policies in online communities are located~\cite{schaffner2024community}, we observed that those policies for GAI online tools are scattered across various policy and support pages. This might be due to, in our understanding, the lack of standardization of how content moderation policies should be presented for GAI online tools and online communities. Below, we briefly describe the common locations of content moderation policies in GAI online tools.

%\mc{say if this is similar or different to the schaffner results and why it might be } \mc{also again think we need to be clear up front in the paper what we consider content moderation - is it removal of content, banning of accounts, etc, need to cite a definition - check brennans paper for what we did there}

The most common page for all tools that includes content moderation policies is the ToS, except Gemini which did not have a separate ToS page. Beyond ToS, 5/14 tools have a separate policy page for \textit{Acceptable Use Policy} (AUP, also named alternatively in some tools, such as \textit{Prohibited Usage Policy} or \textit{Usage Policy}) defining allowed and forbidden usage of the tool, including the criteria of content moderation. For the five tools provided by companies developing multiple products, the ToS of four companies includes policies applicable to content moderation in GAI online tools, and two of these also have AI/GAI product-specific rules regulating content generation activities. Other policy pages we found content moderation policies in are \textit{Service Terms} (ChatGPT), \textit{Community Guidelines} (Midjourney), and \textit{Safety Center} (CivitAI). We also found 9/14 tools have information on content moderation rules in their support pages. In addition to these nine tools, four other tools have support pages but do not contain content moderation policies.

\subsection{Limitations}
Our policy analysis study has several limitations. Our tool list is only representative of commonly used GAI online tools in the US context. Since the market of GAI online tools is monopolized, most customers use only a few types of tools~\cite{forbes_chatgpt_ai_tool_sector, writerbuddy_ai_industry_analysis}, and we cover most of them. Moreover, our policy collection process relied on manual checking on specific pages, meaning we might have missed pages with information on content moderation policies. 

%Moreover, it is worth noting that policy statements could be misaligned with the service provider's practice in real~\cite{10.1145/3512965}, which is a perspective we did not fully investigate in this study.
\section{Study 1: Policy Analysis Findings}
\label{sec:findings1}

\begin{table*}[h]
    \centering
    \renewcommand\arraystretch{1.02}
    \resizebox{1.9\columnwidth}{!}{
    \begin{tabular}{p{0.2\textwidth}p{0.42\textwidth}p{0.42\textwidth}}
    \toprule
    \textbf{Moderation Stages}     &\textbf{Similarity in Policies of Online Communities and GAI Online Tools} & \textbf{Uniqueness in Policies of GAI Online Tools}\\
    \midrule
    Moderation Criteria     &Both online communities and GAI online tools detail a wide prohibition on different problematic content (\cref{subsec: outputtype}) & GAI online tools regulate input, output, as well as output distribution and secondary use (\cref{subsec: criteriapresent})\\
    \midrule
    Moderation Methodology     &Both online communities and GAI online tools describe their moderation systems as a mixture of automatic detections and human reviews (\cref{subsec:toolmethod}) \newline Both online communities and GAI online tools acknowledge their content moderation is not infallible and place responsibility for content on users (\cref{subsec:toolmethod}) & GAI online tools enforce content flagging on both user input and GAI output (\cref{subsec:toolmethod}) \newline Some GAI online tools mentioned safety measures in model training as a moderation method (\cref{subsec:toolmethod}) \newline Only a few GAI online tools provided methods for user-driven content moderation (\cref{subsec:usermethod})\\
    \midrule
    Moderation Consequence     & Both online communities and GAI online tools take down problematic content and punish users (\cref{subsec: consequence_contentuser}) \newline Both online communities and GAI online tools leave users with few options once they have been moderated (\cref{subsec: userappeal}) & GAI online tools take actions on both user input and GAI output, with the goal of preventing the generation and presentation of output (\cref{subsec: consequence_contentuser}) \newline Except legal violations, moderation enforcements on content in GAI online tools are the same across different types of content policy violations (\cref{subsec: consequence_contentuser})\\
    \bottomrule
    \end{tabular}}
    \caption{Major similarities and differences between content moderation policies in online communities and GAI online tools.}
    \label{tab:policycomparision}
\end{table*}

%\mc{definitely we want to structure this section to answer the RQ of how comprehensive were the policies and how do they compare to platforms with user-generated content so that the novelty is clear} 
% \mc{how do these map to the sections in the schaffner paper - more novel would be to say which things are present in both, which are absent etc}
% \lan{We didn't analyze `why service providers moderate content' as Brennan did (referred to as `policy justification' in that paper), since I found the policy justification of GAI online tools more of target the whole policy page (like ToS), and regard AI safety/user safety/responsible use of tool, rather than why they moderate content. Not sure how to elaborate this}

In this section, we present findings about what and how complete the content moderation policies of GAI online tools are (RQ1), given that these tools are still evolving when compared with online communities whose policies have been established for much longer and cover various topics~\cite{schaffner2024community}. We observed that policies in GAI online tools, like those in online communities, cover the three major components of content moderation practice outlined by Singhal et al. \cite{singhal2023sok}: content moderation criteria --- providing definitions of forbidden content (\cref{sec:criteria}); content moderation methodology --- explaining safeguarding approaches to detect problematic content generation (\cref{sec:execution}); and consequences of content moderation enforcement --- detailing the platform responses to problematic content generation and corresponding mechanisms for user appeal (\cref{sec:consequence}). However, we did not find clear presentations in GAI online tool policies of `why content is moderated' which online communities specify --- justifications on why GAI online tools engage in content moderation, beyond general statements on how platforms value user safety and responsible AI development. 
% \chenhao{is the final point in the table?}
% \mc{i wonder if we should say what the differences are earlier? like move this part to related work or the introduction even? seems less like a finding and more just like something we know before we even started}

%\mc{do we define these acronyms somwhere?}\lan{See methodology}

Overall, both online communities and GAI online tools specify similar rules regarding what is not allowed and their moderation strategies at each stage, with GAI online tools modeling their content moderation policies on the relatively mature content moderation frameworks used in online communities. 
%Evidence of this can be seen in the moderation systems implemented in GAI online tools, such as the content moderation guardrails in ChatGPT, which function similarly to content detection systems used in social media -- both rely on classifiers to categorize content into different harm categories~\cite{mahomed2024auditing, openai_moderation_guide}. 
However, online communities have policies that cover user-generated content, while GAI online tools' policies cover user input to the tool (referred to as \textit{input} or \textit{prompt} from here on) in addition to content generated by the tool (referred to as \textit{output} from here on). In most online communities, user-generated content is posted or uploaded directly. Instead, users have less control over the randomness of how the GAI online tools generate output with their input, for which users may unintentionally generate undesired output~\cite{10.1145/3548606.3560599, gehman2020realtoxicityprompts}. Table \ref{tab:policycomparision} summarizes major policy similarities and differences in the two types of platforms. Next, we present details of how GAI online tool policies describe the three content moderation stages.

\subsection{Content Moderation Criteria}
\label{sec:criteria}
\subsubsection{How are Moderation Criteria Specified?}
\label{subsec: criteriapresent}
Similar to how online communities describe prohibited user-generated content in community guidelines~\cite{10.1145/3406865.3418312}, content moderation criteria in GAI online tools are mostly described by articulating what behaviors are acceptable or prohibited when using the service, in the separate AUP, acceptable use guidelines in ToS, and support pages. Interestingly, none of the tools except CivitAI has a separate \textit{Content Policy} or related sections in policy and support pages that specify all rules about forbidden content. 
In short, the GAI online tool policies are often unstructured, and as such, we found content moderation criteria are presented in a mixed and complicated manner. For example, some rules apply across multiple aspects, including input, output, output distribution, and, in some cases, other content hosted within the tool's corresponding services.  

\vspace{-1.2em}
\paragraph{Policies Implicitly Regulate Content Generation Requests.} For all 14/14 tools we studied, each has several acceptable use guidelines using descriptive language to broadly define acceptable or prohibited tool usage, usually starting with \textit{``Do not use the service to ...''}. These guidelines apply to all user behaviors within GAI online tools, including when they generate content using the tool.
%Common prohibited tool usages related to content generation requests include engaging or facilitating illegal activities (x/14), self-harming (x/14), harming others (x/14), deceiving others (x/14), violating other's rights (x/14), and using for professional purposes (e.g., health, law) (x/14). 
Notably, all 14 tools we studied forbid users from overcoming system restrictions (or `jailbreaking' the system), which implicitly rules input. 

\vspace{-1.2em}
\paragraph{Policies Explicitly Target Content.} In addition to descriptive acceptable use guidelines, each GAI online tool lays out rules on what content is prohibited that are scattered over policy and support pages, much like in online community content moderation policies~\cite{schaffner2024community}. GAI online tools typically talk about `content' which encompasses both input and output, represented by the following definition: \textit{``You may provide input to the Services (`Input'), and receive output from the Services based on the Input (`Output'). Input and Output are collectively `Content'.''} (ChatGPT's ToS). Many rules solely use the term `content' when describing prohibitions, and they do not distinguish if there are different rules for different modalities of content (e.g., text or image) or depending on different tasks (e.g., dialogue versus creative writing). %A few GAI online tools just make a set of unspecified rules and apply them to both input and output generally: \textit{``When using our onsite image generating services, \textbf{all prompts and resulting images must comply with our content policies.}''} (CivitAI's \textit{Safety Center}). 

Furthermore, some image generation tools are integrated into or built up with online communities for sharing user-created AIGC (i.e., Firefly, Craiyon, CivitAI, and Midjourney). We found that policies there just describe rules about any content users have as input/output in GAI online tools and uploaded content in online communities solely as `content'. As Midjourney defines `content' in its ToS: \textit{``Inputs, Assets, and other content such as messages, photos, videos, and documents that you may provide to the Services (such as through uploading, posting, sharing, or chat messages) are collectively, `Content'.''} (Midjourney's ToS) Therefore, it is sometimes unclear whether a rule applies to all or specific types of content defined by these tools. %This also makes it hard for users to know if there are additional rules depending on whether the content is input/output in a GAI online tool, or a post/comment in an online community associated with the tool.

%or whether a rule applies to input/output or user-generated content in an online community associated with a GAI online tool.
\vspace{-1.2em}
\paragraph{Prohibitions on Output, Input, and Output Distribution.} Rules clearly governing output often start with \textit{``Do not generate/create content that ...''} if without directly stating the term `output'. We noticed that all tools we studied except NovelAI (13/14 tools) elaborate on what output is restricted by defining rules in their policies. 
%For example: \textit{``[Do not] Generate deceptive or misleading digital content such as fake reviews, comments, or media.''} (Claude's \textit{Usage Policy}). 
%\mc{should we discuss input first, then output rules, seems more logical, i would swap these} 
Meanwhile, 10/14 tools included specific rules outlining criteria for input. Besides restricting input that leads to prohibited output, these rules mostly outline the same restrictions as rules targeting output, with many defining forbidden input along with forbidden output in the same place. For example: \textit{``Do not create images or use text prompts that are inherently disrespectful, aggressive, or otherwise abusive.''} (Midjourney's \textit{Community Guideline}). %We further elaborate on what type of input and output is prohibited in \cref{subsec: outputtype}. \mc{why do we elaborate in a different section?}

%Notably, input-specific rules tend to focus more on prompts that violate other's rights like privacy, as mentioned by x/14 tools. 
%
% \mc{this following part seems unique to GAI tools?}
% \lan{I would like not to underline the uniqueness of this part compared to online communities, since it's too apparent..}
We also found that 12/14 GAI online tools specify rules that address the future distribution and secondary use of output, despite the limited practical authority they have to enforce these rules. Similar to rules governing input, rules on output distribution often reiterate the same restrictions as those applied to output and, in many cases, are presented alongside the output restrictions: \textit{``[Do not] Creating, generating or distributing content that depicts gratuitous violence, cruelty, abuse, sex or gore.''} (You.com's AUP). Additionally, we noticed special restrictions that applied to output distribution only. 8/14 tools forbid users to mislead others that the output from the GAI online tool is created by humans: \textit{``[You will not] Represent any Output (defined below) as human generated when they are not.''} (xAI's ToS). 5/14 tools prohibit users from using the output to develop machine-learning models: \textit{``[You may not] Use Output to develop models that compete with OpenAI.''} (ChatGPT's ToS).

\vspace{-0.3em}
\subsubsection{What Types of Content Are Forbidden?}
\label{subsec: outputtype}
In addition to the loose and scattered structure, content moderation criteria are articulated using a variety of terms that cover a broad range of restrictions, similar to how community guidelines in online communities outline content prohibitions~\cite{10.1145/3406865.3418312}. This observation on the varied use of terms in moderation criteria also aligns with findings on AUP of foundation models~\cite{klyman2024acceptable}, where more than 120 prohibited behaviors are described using diverse terms. However, we noticed that forbidden content defined in the GAI online tool policies could be grouped into broader categories, based on the nature of each prohibition. Drawing on major content moderation topics in online communities recognized in prior works~\cite{schaffner2024community, gillespie2017governance, singhal2023sok}, we mapped out four types of prohibited content at a high level. The four categories apply to all content span input, output, output distribution, and include rules that do not specify the type of content: \textbf{harmful content} that harms individuals, groups, and public safety, regardless of legal or illegal; \textbf{content that violates other's rights} with privacy and intellectual property/copyright violations; \textbf{misleading content} such as mis/disinformation and deceptive content, including content with misleading nature; and \textbf{content that is not appropriate for everyone} which depicts sexual and violence. We found that all tools except NovelAI outline all four types of prohibitions in their policies.

We note that the first three categories are frequently addressed in online community policies, where most mainstream platforms enforce strict prohibitions~\cite{schaffner2024community, singhal2023sok}. In contrast, the restrictions of the fourth category, which are strictly prohibited in many GAI online tools, largely depend on the platform's nature when applied to online communities, where personal content moderation (e.g., personalized filtering) is commonly used instead of direct bans from the platform~\cite{singhal2023sok,jhaver2023personalizing}.

\subsection{Content Moderation Methodology}
\label{sec:execution}
\subsubsection{How is Problematic Output Detected or Prevented?}
\label{subsec:toolmethod}
GAI online tools outline various approaches used for their moderation systems in their policies. These include interventions outside the GAI model for content flagging with similar strategies used in online communities~\cite{schaffner2024community, singhal2023sok}, as well as improving the GAI model itself to generate safer output, which is a GAI-specific strategy. Meanwhile, GAI online tools also acknowledge their lack of ability to moderate output as well as online communities do in moderating user-generated content~\cite{schaffner2024community}, putting all liability of input and output on users.

As with the moderation method disclosed in policies of online communities, GAI online tools outline themselves as employing both automated detectors or filters (7/14 tools) and human reviews (5/14 tools) in their moderation systems. These approaches are enforced in both input and output, represented in the following policy: \textit{``Your prompts and the results generated [...] may be reviewed through both automated (e.g., machine learning) and manual methods for abuse prevention and content filtering purposes.''} (Firefly’s \textit{User Guidelines}). Nevertheless, we noticed that many tools tend to be vague on technical details and nuances of these interventions. For example, another policy in Firefly’s \textit{User Guidelines} only explains its output flagging as \textit{``use available technologies, vendors, or processes''}. %\lan{Nobody mentions that, but the proportion of automatic detection and human reviews are definitely different between online communities and GAI}

Simultaneously, we found five tools driven by self-developed foundation models that claim themselves dedicated to training and improving their models for output safety. As exemplified by the following instance: \textit{``We also work to make our models safer and more useful, by training them to refuse harmful instructions and reduce their tendency to produce harmful content.''} (ChatGPT’s \textit{Usage Policy}). Occasionally, policies refer readers to model documentation for further details of these approaches: \textit{``Limitations and bias in AI are still being researched and we're working actively on this subject. You can learn more in the DALL·E mini model card.''} (Craiyon's FAQ section on the home page).

% The next paragraph is optional and can be removed considering the page limits
%Notably, in contrast to the fair amount of reports on safeguards that target the content creation activities and output, we found few tools present how they take precautions on malicious output distribution and use. The only exemption is Adobe Firefly which integrates credential watermarks in the output images, disclosed in its policy as below: \textit{Adobe is committed to building trust and transparency into digital content with Content Credentials [...] \textbf{``Adobe automatically attaches Content Credentials to images created in Firefly to show that they were AI-generated.}''} (Adobe Firefly's Question section on the home page). 

Meanwhile, 13/14 tools have disclaimers when it comes to who is responsible for the output generated using the GAI online tool and its moderation. In these tools, these disclaimers about the guaranteed safety of generated output frequently cite the well-known unpredictability of GAI output~\cite{10.1145/3548606.3560599, gehman2020realtoxicityprompts}. As illustrated by the following policy: \textit{``This use of Al is relatively new and still evolving. As a result, while we have taken - and continue to take - efforts to preclude your creation of extreme content, we cannot guarantee the suitability or appropriateness of the resulting images you generate.''} (DreamStudio’s ToS). Some tools also acknowledge the potential failure of their safeguards, similar to the online community acknowledgments of challenges in moderating user-generated content: \textit{``These [safety] features are not failsafe, and we may make mistakes through false positives or false negatives.''} (Claude's \textit{Help Center}). 12/14 tools say users are ultimately liable for the input, output, and output distribution and use of GAI online tools. For instance: \textit{``You are solely responsible for your Input [...] You are solely responsible for the creation and use of the Output and for ensuring the Output complies with the Terms.''} (Firefly’s \textit{ToS}).
 % \mc{add a takeway for each section - what is the main point(s) about the findings in this section we should leave the section knowing}
\subsubsection{How Can Users Combat Problematic Output?}
\label{subsec:usermethod}
In addition to putting users in charge of their input and output, service providers sometimes allow users to engage in the moderation process by reporting problematic output to them. Some reporting channels specifically target the copyright infringement of output, as mentioned by 6/14 tools. Similar to what was observed in online communities~\cite{schaffner2024community}, GAI online tool policies usually claim a unified reporting pipeline for copyrighted content, grounded by existing laws (i.e., Digital Millennium Copyright Act, DMCA), and to fulfill legal requirements.

%Most of the report channels specifically target the copyright infringement of output, raised by policies of x/14 tools. They usually claim a unified report pipeline of mailing a hand-written notification to the service provider's legal agent, with a few exemptions also offering online portals for this:\textit{``If you believe that your intellectual property rights have been infringed, please \textbf{send notice to the address below} or \textbf{fill out this form}.''} (ChatGPT's ToS). As claimed by some tools, setting up this type of channel is grounded by existing laws (i.e., DMCA) and, as we suspected, more of fulfilling legal requirements. The prevalence and standardized copyright infringement report channels were also observed in content moderation policies of online communities~\cite{schaffner2024community}. \lan{considering cut off this paragraph to only 1-2 sentences}

Beyond copyright infringement reports, we only found 5/14 tools mentioning how users can combat problematic output in general. This is surprisingly contrary to online platforms, which heavily rely on user-driven moderation even beyond copyright infringement~\cite{schaffner2024community, seering2020reconsidering}. Although channels for user engagement in content moderation are limited in GAI online tools, some tools provide feedback mechanisms integrated into the tool interface (e.g., functions to instantly `thumb-up' and `thumb-down' output) and direct contacts with the service team (e.g., report forms or contact email). As summarized by this example quote: \textit{``Users can report problematic or illegal content via the Feedback button or the Report a Concern function.''} (Copilot's ToS).

\subsection{Content Moderation Consequence}
\label{sec:consequence}
\subsubsection{What Happens to Problematic Content and Users?}
\label{subsec: consequence_contentuser}
GAI online tool policies elaborate on how they respond to problematic input and output, as well as users engaging in problematic output, similar to rules disclosed in policies of online communities~\cite{schaffner2024community}. As summarized by the following policy: \textit{``Content that violates our rules, or attempts to circumvent our content restrictions, will result in appropriate actions, which may include content removal, flagging of the account, suspension of access to the image generation feature, or a ban from the platform.''}(CivitAI's \textit{Safety Center}) 

User-targeted moderation responses in GAI online tools are nearly the same as those in online communities. Although most GAI online tools say account restrictions target general term violations or can be enforced for any reason, we found 5/14 tools relate these account-level actions to problematic content generation specifically. %Additionally, account-level punishments are usually enforced on users who break content policy seriously, such as those who repeatedly create problematic content or engage in illegal content. 
For example: \textit{``We have adopted a policy of terminating, in appropriate circumstances, Users who are deemed to be repeat infringers [of copyright].''} (xAI's ToS). 3/14 tools also issue warnings to certain accounts with users prompting for content that violates the content rules, exemplified by the following policy: \textit{``As part of our safety process, we warn users if we believe their prompts are violating our Usage Policy.''}(Claude's \textit{Help Center}).

In comparison, content-targeted responses (10/14 tools) share a similar strategy with those in online communities but have a different goal. Instead of focusing solely on removing problematic content like in online communities, GAI online tool policies aim to refuse the GAI processing of problematic input and prevent the presentation of undesirable output. Thus, flagged input is typically blocked or removed from the system, with an error or warning message returned instead of the requested output. If problematic output is generated, it may be removed or blurred afterward. As represented by the following quote: \textit{``Image Creator may block prompts that violate the Code of Conduct, or that are likely to lead to the creation of material that violates the Code of Conduct. Prompts or Creations that violate the Code of Conduct may be removed.''}(Copilot's ToS of Image Creator). We found no distinction in these enforcements across policy violations, unlike online communities which, for example, investigate harmful speech and misinformation before content removal but remove copyright violation content immediately~\cite{schaffner2024community, singhal2023sok}. One exception is certain legal violations, where service providers may take additional legal action on both input and output. For instance: \textit{``We report apparent child sexual abuse material (CSAM) to the National Center for Missing and Exploited Children.''}(ChatGPT’s \textit{Usage Policy}).

\subsubsection{What Users Can Do After Being Moderated?}
\label{subsec: userappeal}
Similar to policies of online communities~\cite{schaffner2024community}, we noticed policies of GAI online tools give users few options after they get moderated. While most of the tools have terms for legal disputes, user appeal after general content moderation is not well specified in the current GAI online tool policies for the tools we examined -- only 5/14 tools we studied clearly outline non-legal appeals users can engage in after getting moderated. Furthermore, most appeals are only applicable to redress account restrictions, rather than content-targeted enforcement like blocked input and removed output. 

Beyond simple user appeals, there is still little that can be done after content moderation. We observed that none of the tools except three talk about user feedback if they think any content-targeted responses of moderation are questionable. We even found an extreme case, where the service provider does not offer an appeal or ask for any feedback but asks users to try again if the output gets moderated: \textit{``Q: Why are some of my images blurred? [...] [T]he model will blur out any content that may be considered inappropriate or offensive. [...] If you are unhappy with the results, you can always try again with a different prompt.''} (DreamStudio’s FAQ).

%\chenhao{I think having a big table that summarizes these results would be great. It reads very scattered and it is unclear what I should take away.}
\section{Study 2: Reddit Study Methodology}

To answer our second research question, we conducted a case study focusing on content moderation when users engage in AIGC creative tasks such as creating fiction and art. To do so, we analyzed Reddit posts on discussions about content moderation experiences in AIGC creative tasks using GAI online tools. Our approach follows one that is commonly used in prior works to gain real-time insights from people, such as understanding user perceptions and reactions after content moderation in online platforms \cite{ma2021advertiser, kou2024community}, by qualitatively analyzing online discussions.
%\mc{do we have any other cites to add?}
We performed a keyword-based search in GAI online tool-related subreddits and then manually filtered out irrelevant posts, creating the final Reddit dataset for analysis. Finally, we performed qualitative analysis on randomly selected posts and corresponding comments from our dataset. 
%In the following section, we present our procedure in detail.

\subsection{Reddit Post Collection}
We performed Reddit post collections in October 2024 through keyword-based searching. We started by deciding which subreddits to focus on and creating a keyword list of content-moderation-related words. Next, we scraped Reddit posts through the Python Reddit API Wrapper (PRAW),\footnote{https://praw.readthedocs.io/en/stable/} and did another round of manual filtering to get the posts we wanted. 

\textbf{Subreddits Choice.} Our selection of subreddits for data collection follows two criteria. First, we only included subreddits that solely discuss tools in our list from Study 1 (\cref{sec:toolist}), to exclude discussions on content moderation beyond GAI online tools. For example, most posts discussed AIGC creation in r/aiArt, but we did not include this subreddit since it was difficult to determine if a discussion was about using GAI online tools. We also only considered subreddits within the top 5\% of all subreddits, to collect high-quality discussions from active communities. Complying with the two criteria above, we finally selected seven subreddits to perform data collection (ordered in size): r/chatgpt, r/midjourney, r/dalle2, %(for Dall-E in both ChatGPT and Microsoft Copilot)
r/claudeAI, r/Bard, r/dalle, and r/perplexity\_ai.\footnote{We excluded r/NovelAI since we found NovelAI has enforced almost no content moderation, based on Study 1 analysis and discussions in this subreddit.} Discussions on these subreddits span six GAI online tools: ChatGPT, Claude, Copilot (DALL-E image generator only), Gemini, Midjourney, and Perplexity AI.

\textbf{Keyword List Creation.} Referring to prior works on analyzing public discussions of online platform content moderation~\cite{ma2021advertiser, kou2024community}, we picked 12 words that were frequently used when talking about content moderation: `moderate', `censor', `ban', `block', `suspend', `restrict', `warn', `flag', `appeal', `violate', `terminate', and `remove'. Using these keywords, we first performed an open search in r/chatgpt and r/dalle2 through the Reddit website, where two researchers reviewed 25 posts per subreddit related to content moderation in AIGC creative tasks, to get a sense of the common discussion themes. Four additional keywords frequently used in those posts were identified: `content policy', `guardrail', `filter', and `refuse'. These collective keywords made up the final keyword list that we utilized for data collection.

\textbf{Scraping Posts.} We searched and recorded posts via PRAW with either a title or selftext (content of the original post) containing at least one of the keywords in our list. We used all forms and tenses of the keywords in this process. For example, when matching for the keyword `censor', we used `censor', `censored', `censoring', and `censorship'. We did not apply time constraints in searching, indicating that all posts from when the subreddits were established to October 2024 (the time of data collection) were under the search scope. We collected a total of 5185 posts through this process.

\textbf{Dataset Cleaning.} We noticed a high false-positive rate on our collected posts, due to the broad scope and context of keywords used beyond content moderation in AIGC creative tasks. To reduce the high false-positive rate, we manually checked all posts 
%\mc{who is we, how were they checked?} 
and filtered out the irrelevant ones. To be specific, three researchers checked the selftext of each post, only keeping those that discussed a general opinion on content moderation in GAI online tools, or a user experience of being moderated when engaging in AIGC creative tasks using GAI online tools. After that, we utilized PRAW to get all corresponding comments on the remaining posts. We cleaned up the comments by removing those shown as `[deleted]' or `[removed]'. Through this process, we retained 1123 posts and 33465 corresponding comments, constructing our Reddit dataset.

% For example, the dataset initially contained many posts discussing content moderation of AIGC in social media, jailbreaking, or completely irrelevant topics through the keyword-based searching process. 

\subsection{Data Analysis}
Due to the high volume of posts and comments, we analyzed a portion of the dataset rather than the whole dataset, following the common practice in prior works conducting qualitative analysis on Reddit (e.g., \cite{ma2021advertiser,299651,kou2024community}). We randomly selected 130 posts from our dataset, which corresponded to 3839 comments, for our data analysis. All posts and corresponding comments were imported into MAXQDA for further analysis. The statistics of the final Reddit dataset and sampled posts for data analysis are shown in Table \ref{tab:reddit} in the Appendix.

We performed an iterative, deductive coding and thematic analysis. First, we ran another random sampling to split our sampled dataset into 30 posts (971 comments) and the other 100 posts (2868 comments). The first author then reviewed and performed an initial coding on the set of 30 posts and corresponding comments to build the codebook. During this process, the research team held regular meetings to review sampled posts, discuss questions, and refine the codebook. 

All 130 posts and corresponding comments were then coded iteratively by three researchers using the codebook. First, all posts were divided equally into three sets with a comparable number of comments and assigned to each coder for primary coding. These sets were later reassigned among the coders for secondary coding, ensuring that each post and comment was coded by at least two coders. Three coders met regularly to discuss the coding progress, compare each other's codes, and solve discrepancies during secondary coding. We did not calculate IRR, as the iterative analysis was informed by our data and all disagreements were resolved during the process~\cite{10.1145/3359174, armstrong1997place}. 

We reached thematic saturation midway through analyzing these 130 posts---where no new codes emerged---indicating that our analysis was comprehensive for extracting qualitative findings~\cite{saunders2018saturation}. Thus, we did not sample additional posts.
%\mc{why didn't we calculate IRR? Cite McDonald paper and explain, do the other reddit studies calculate IRR}

\subsection{Limitations and Ethics}
We acknowledge a few limitations of our Reddit study. First, our data collection only covered discussions on six GAI online tools. We also observed an uneven distribution of Reddit discussions across different tools. For instance, while discussions on content moderation in ChatGPT are extensive, we found only seven related posts about Perplexity AI. Additionally, our analysis was limited to a sample set of user perceptions posted in public discussions. Therefore, without a broader analysis, our study may not have assessed all successes and failures around content moderation policy enforcement. 

Although all posts and comments we collected and analyzed are publicly accessible, there are potential privacy violations for the users who post them, a widely acknowledged ethical concern of social media and Reddit research~\cite{fiesler2024remember}. Therefore, when reporting any post or comment in the paper, we removed all identifying information and adjusted some wording into synonyms to prevent a direct search. Moreover, the Reddit study was reviewed by our Institutional Review Board (IRB) before the data collection. 

%Our approaches to minimizing ethical concerns follow the common practices in prior qualitative Reddit research (e.g.,~\cite{299651, ma2021advertiser}).

\section{Study 2: Reddit Study Findings}
\label{sec:findings2}
In this section, we present findings on how content moderation policy enforcement in GAI online tools succeeds and fails (RQ2), from the case of user experience of using these tools for AIGC creative tasks. 

%\mc{is this only for creativity tasks or did we just study all posts in the gAI tools subreddits, since that is misleading otherwise, but correct if i am wrong, if it is only creation, why does the first findings section not cover anything related to the task type or creation? or creativity}
%\mc{need to rework this intro paragraph so its more clearly tied to study 1}  
%We start by introducing the user's overall experience of content moderation -- types of content moderation users confront and their behavior after being moderated (Section \ref{sec:overallexperience}). Then, we elaborate on users' criticism of content moderation they experienced in AIGC creation processes \mc{again, lets either define this upfront in the paper or be more specific about what we mean by AIGC creation processes} (Section \ref{sec:badexperience}), as well as their positive opinions towards the current deployment of content moderation (Section \ref{sec:goodexperience}).

\subsection{Successes of Moderation System: Blocking Malicious AIGC Creation Attempts}
\label{subsec:success}
Based on user-shared examples in public Reddit discussions we analyzed, we found that moderation systems in GAI online tools are generally effective in detecting and blocking AIGC creations and requests that maliciously violate moderation criteria, such as requests for or generated content involving pornography, scams, or hate speech. We observed instances where users posted about how problematic AIGC creations were either denied at the request stage or immediately removed after generation across all six GAI online tools. 

We found users posted about their satisfaction and appreciation with current content moderation enforcement for 3/6 tools, regarding its success in blocking malicious attempts. %Some users criticize complaints about content moderation from malicious users requesting problematic AIGC, asserting that such requests should unquestionably be moderated. As a ChatGPT user commented under a post discussing many content moderation examples in AIGC creative tasks: \textit{``I use it for productivity and not for meaningless tasks or jokes. Lots of posts I've been seeing are along the lines of `it won't mimic this controversial figure or come up with atrocities.' I don't know what else to say.''} Simultaneously, 
Some users strongly recognized the potential harms of malicious AIGC and, therefore, appreciated the comprehensive moderation criteria and moderation systems that GAI online tools employ to successfully protect individuals and society from harmful content. This is exemplified by a user who commented on a request for porn to be moderated in ChatGPT's DALL-E: \textit{``Basically no deep fake images that could be used to falsify evidence of adultery [are allowed]. I'm really glad this sort of thing wasn’t available in my teens, and that [OpenAI] is protecting people from it now.''}

%\mc{findings 1 and 2 are not linked if the findings 2 is creative tasks only - can we adjust findings one or add something to also bring up creative tasks sooner}

\subsection{Failures of Moderation Pipelines: Moderation System and User Support Failures}
Despite evidence that moderation systems effectively block malicious AIGC creation, we found that many users discussed failures in the current moderation practices on GAI online tools when performing AIGC creative tasks. These cases span not only the moderation systems themselves that failed to make justified moderation decisions (\cref{subsec:systemfail}), but also the post-moderation stages, where service providers failed to assist users in understanding and redressing moderation decisions (\cref{subsec:providerfail}). Although some user-reported moderation failures here resemble those in online communities (e.g., biased and inconsistent moderation in social media~\cite{lyons2022s, vaccaro2020end, haimson2021disproportionate}), their impact on user experience extends beyond the frustration recognized in moderation failures within online communities \cite{myers2018censored} -- moderation failures in GAI online tools have hampered the usability and capability of GAI, \textit{``making [the tool] completely useless''} for users. Next, we expand on how users perceived failure cases of the moderation system and user support, and how they blocked the usability of GAI online tools for AIGC creativity tasks.

\subsubsection{Failures in Moderation System Decision-Making}
\label{subsec:systemfail}
According to public Reddit discussions we examined, users frequently talked about how the moderation systems, especially the algorithmic and automatic ones, behaved inaccurately, inconsistently, and unreasonably when GAI online tools were used for AIGC creative tasks. While similar failure cases have been shown by researchers through model testing and auditing (e.g., \cite{mahomed2024auditing}), we observed that users developed their own `folk theories' --- collective understanding of system operation based on personal knowledge~\cite{eslami2016first} --- on how and why moderation systems fail in real use cases. 

\vspace{-1em}
\paragraph{Failure in Mitigating False-Positive Rate (5/6 tools).} When discussing the moderation system inaccuracy, users frequently posted the pervasive false-positive moderation decisions -- where users were moderated for using harmless prompts to generate AIGC that was not intended to violate any rules. Users discussed that some false positives were just random bans issued by moderation systems~\cite{gomez2024algorithmic}, or because their prompts resulted in arbitrary malicious outputs due to the randomness of GAI output~\cite{10.1145/3548606.3560599, gehman2020realtoxicityprompts}. As a comment reasoning a random false-positive moderation case in Copilot's DALL-E: \textit{``That [harmless input] still allows for random imagery that you aren't in control of. You've input token words, but whatever is composed might break the filters. It's just an RNG game with loosely curated results.''}

Meanwhile, users also reported that some false positive decisions to stop a request from being processed might be owing to the over-sensitive input moderation. Users observed a high likelihood of requests blocked if they were using input with, or requesting content correlated to moderation criteria, even if their requests were not to create AIGC that violated policy. For example, requesting images of children could be falsely flagged as child abuse. Furthermore, many posts mentioned the over-sensitivity of banning inputs containing certain words. This input moderation strategy, although not explicitly detailed in any policy because of its ambiguity in disclosing details of moderation methods (\cref{subsec:toolmethod}), was widely recognized by users as being implemented in image generation tools such as Midjourney and DALL-E. A ChatGPT's DALL-E user wrote: \textit{``When asking for a portrait of a person, `headshot' is banned. Of course, it is for violent reasons. But double meanings in English abound.''} Table \ref{tab:overreaction} in the Appendix presents examples of false-positive moderation on input related to each moderation criterion. %\mc{might be good to summarize some of the false positives here} \lan{will leave the comment here and add quotes after trimming down}

Policies acknowledged that there could be false-positive moderation (\cref{subsec:toolmethod}), and users spoke of circumventing these cases through tricks like jailbreaking~\cite{298254, jin2024jailbreaking}. Yet, many users argued that the frequency of false-positive moderation in practice was excessively high, which impaired normal tool usage and frustrated users. A former ChatGPT's DALL-E user shared their experience: \textit{``I got too many warnings for nothing [...] I just find myself thinking of something to try, then becoming afraid of triggering arbitrary warnings, and then just not trying anything. It's resulted in me just not using it.''} 

\vspace{-1em}
\paragraph{Failure in Making Moderation Decisions Consistently (5/6 tools).} Users also discussed the inconsistency of moderation systems -- where moderation systems' behaviors varied across different users or responded differently to similar or identical requests that should uniformly be either moderated or allowed. For example, a Copilot’s DALL-E user questioned the moderation system when two requests to create copyrighted content produced different outcomes, where one was processed while the other was moderated: \textit{``I tried `Batman and Catwoman getting married' and got caught by the filter. Somehow `Nightwing and Starfire getting married' worked.''} %The inconsistency of moderation systems could also represented by arbitrary moderation decisions, as a Midjourney user summarized: \textit{``an identical same prompt will get through sometimes and banned other times''}. 

Most inconsistencies could be attributed to the inherent bias of moderation algorithms~\cite{mahomed2024auditing, binns2017like}, as well as the randomness of algorithmic moderation systems and GAI output, as some users also realized. Some other users attributed this to intentional unfair enforcement in moderation, speculating that undisclosed implicit or non-transparent rules existed in GAI online tools. For example, a ChatGPT user speculated that the moderation systems work differently in different accounts based on prior user behavior: \textit{``Sometimes I think ChatGPT works differently based on previous interactions. It always quit with my friends who constantly try to break it, but it let 95\% of my attempts pass.''} Some users, however, expressed distrust of service providers regarding potential implicit or non-transparent rules. A ChatGPT's DALL-E user criticized: \textit{``I tried to use a prompt with Elon Musk's name in it. That’s not allowed. Neither is Joe Rogan [...] But I’m allowed to use other high-profile people's names? Seems like the Developers are letting their personal bias get in the way.''}
%\mc{each subsection needs a takeaway sentence - what should someone remember after reading the section?}
%the inconsistency in moderating content related to famous people, describing it as a biased rule deliberately implemented by the service provider

%The inconsistent moderation is also evident in the unequal enforcement of moderation criteria across different AIGC creation requests. A Copilot's DALL-E user, for example, questioned the moderation system that two requests for creating copyrighted content yielded different outcomes, with one processed while the other got moderated: \textit{``I tried `Batman and Catwoman getting married' and got caught by the filter. Somehow `Nightwing and Starfire getting married' worked.''} A user shared a potential example of the political bias of writing moderation in Gemini: \textit{``I wanted Gemini to write a poem in the voice of Trump. It wouldn't do it, saying I should Google for updated news and politics. It had no issue with writing a poem in the voice of Biden.''} 

% Based on the posts we examined, many users have posted about these inconsistencies and seem to have eroded user trust in the stated policies, even raising speculation about the existence of implicit rules. \lan{add quote}

%\mc{make sure you write the findings in a way that grounds it in the data...use words like users posted about, in the dataset we found posts that...etc to make it clear where the findings arose from}
\vspace{-1em}
\paragraph{Failure in Enforcing Moderation Criteria Regarding the Context of Tasks (6/6 tools).} Researchers have found that GAI content moderation guardrails overly-censored cultural content through algorithmic auditing~\cite{mahomed2024auditing, riccio2024exploring}. This finding aligns with real user experiences -- many discussions we investigated highlighted frequent and extensive restrictions in doing AIGC creative tasks. Although many moderation cases broadly aligned with the moderation criteria (\cref{subsec: outputtype}), users argued that these decisions should be reconsidered in the context of AIGC creative tasks, feeling they were being overly restricted.
%as they felt these tasks were overly restricted by the moderation system.

Users frequently reported moderation examples when they were creating fictional, horror, and fantasy writing and art, with prompts and AIGC creations being flagged as harmful or violent. Another type of content users reported frequently being prohibited was romance and artistic content, which was classified as sexual. We also found various and scattered discussions on the requests being blocked or the AIGC creations being removed, spanning requests for historical materials, autobiographies, educational materials, propaganda, and creations involving jokes, sarcasm, or profanity. Users also perceived divergences from their requests in writing tasks as over-restrictive moderation practices. They reported that the generated writing was often rendered plain or positive, failing to write debatable topics even without policy violations. We note that this phenomenon mainly arose from value alignment of LLMs --- a process of fine-tuning models to produce responses that adhere to widely accepted opinions \cite{han2022aligning, gillespie2024generative} --- rather than content moderation mechanisms. Table \ref{tab:overrestriction} in the Appendix lists examples of AIGC creative tasks users tried to complete or completed with harmless intentions, when they felt they were over-restricted under each moderation criterion. %\mc{think we need to give examples here and not just in appendix}

Users argued that the one-size-fits-all moderation enforcement failed to consider user intentions and the intended use of AIGC, thereby risking normal creative processes. A ChatGPT user posted: \textit{``Even if the generated text is about something bad, it could potentially still be used for good. A story set in medieval times will probably contain fighting and violence. The story can still be considered good if people like reading it.''} Users spoke of how the broad definition of problematic content was, in some cases, essential for certain creations like fictional writing and art. As a ChatGPT's DALL-E user argued: \textit{``I agree that depictions of violence and some nudity should be perfectly fine to generate as long as they're not photorealistic or otherwise problematic. After all, both are extremely prevalent in art.''} Moreover, users stated that without awareness of the context of moderation enforcement, the GAI online tools were now more restrictive than most digital and physical platforms. For example, many moderated requests were intended to generate content that, according to users, was \textit{``perfectly acceptable on network TV''}, or even \textit{``permissive in books marketed at early teens''}. 

Meanwhile, users complained about overly restrictive moderation enforcement on AIGC creative tasks because they limit GAI's creativity and generative capabilities. For example, a representative comment by a Gemini user spoke of getting moderated in writing political and historical topics: \textit{``These experiences lead me to believe that Gemini's strict content filtering significantly limits its ability to engage in a wide range of topics, potentially hampering its usefulness.''} %Talking about how content moderation limits ChatGPT's use for creativity when writing fiction, a user wrote: \textit{``The implications go much wider than generative fiction though. It's going to make a lot of the material very bland and one note, whatever thing it is you want to generate.''} 

\subsubsection{Failures in Supporting Users After Moderation}
\label{subsec:providerfail}
In public Reddit discussions we analyzed, users frequently expressed frustration over their experiences after being moderated. They discussed unclear explanations of moderation decisions, ambiguous policies that hamper the reasoning of moderation decisions, and limited assistance provided in user appeals. Failures in supporting users after moderation further undermine the usability of GAI online tools.

\vspace{-1em}
\paragraph{Failure in Providing Clear Explanations of Moderation Decisions (5/6 tools).} Users expressed their confusion about the explanations provided when GAI online tools moderated their input or output. Similar to what happens after content removal in online communities~\cite{myers2018censored}, when users got moderated in a GAI online tool either resulting in system refusal or account ban, they sometimes received only a generic notice of policy violation or experience a plain system-generated refusal like \textit{``I can't assist with that request''}. These responses typically lacked specific details about where, what, and how the policies were violated. 
%as raised in a post by a Midjourney user: \textit{``I suddenly lost access to the service without any clear explanation. The only communication I received was a message stating that I had violated the ToS. However, I am confident that I did not breach any of the outlined terms.''}

In text and multi-modal generation tools, even if users obtain a GAI-generated detailed explanation of moderation decisions, it might further confuse the users not because of the lack of response transparency, but due to the hallucination of GAI output~\cite{zhang2023siren}. As reported by users in ChatGPT, Claude, and Gemini, they could always ask the system to generate the reason their input or output got moderated. However, the explanations generated by GAI were nonsensical. A ChatGPT's DALL-E user shared: \textit{``I had one image out of four hit the censorship today, was doing pencil sketches of foxes. I asked why they were censored, and ChatGPT said it didn't know why.''} Another ChatGPT user reasoned this phenomenon as: \textit{``ChatGPT doesn't actually know why the content policy kicked in. If you ask, it will just make up something based on the prompt. But the real reason isn't actually known by it.''} Unreasonable justifications generated by GAI systems were also observed when the system returned a refusal with an explanation directly. As an example shared by a Gemini user: \textit{``I asked it to write a script for an ASMR video featuring hypnosis and finger snaps and it refused because it could be dangerous and should be done by a professional.''} %This is retarded.

\vspace{-1em}
\paragraph{Failure in Presenting Clear Moderation Criteria in Policies (3/6 tools).} Regardless of the widely acknowledged content moderation criteria in policies (\cref{subsec: outputtype}), users argued that these criteria often lacked details or, in some cases, were not provided at all. Some users questioned the clarity of moderation criteria explanations. For example, a Midjourney user remarked: \textit{``[The explanations are] all kinds of vague. `Avoid nudity but also avoid fixation on the naked breast.' So is the male torso OK or not?''} The problem of unclear criteria was particularly severe in image generation services that blocked input containing certain words, as none of them provided a list of banned words to the public. Many users argued that service providers should address this lack of transparency by presenting the banned word, along with explanations for why each word was prohibited. As a ChatGPT's DALL-E user criticized: \textit{``How as a user am I supposed to know all of the `no' words when I'm trying to edit an image and each bad strike counts towards a hidden `ban' counter? [...] A product should clearly define what is and what is not acceptable through its ToS. If a prompt is `bad', the reason should clearly be spelled out so that a user knows not to pursue that. ''}

%\mc{do we have a citation on how many tools do that? os is that in their policies? do they provide the banned word list?} \lan{see the later part of third paragraph of section 6.2, or maybe we should make it more stand out}

The unclear moderation criteria, along with users receiving what they perceived as inadequate and unreasonable explanations after moderation, left users without a reliable framework to interpret moderation decisions. When triggering moderation, users relied on self-reasoning about how the moderation system made the decision, which policies they might have breached, and if the moderation decision was a system failure. This situation, users argued, severely compromised the tool's usability. A ChatGPT user wrote: \textit{``Now imagine having to redo a message 10 times to find exactly what words trigger the filter. I'm not left with a lot of messages to do my thing.''}

%\mc{better tie these findings to study 1, e.g. appeal processes not well specified and we also noticed that people rarely mention them in the posts we examined}
\vspace{-1em}
\paragraph{Failure in Effectively Supporting User Appeals (3/6 tools).} User appeals were not widely acknowledged in policy (\cref{subsec: userappeal}). Nevertheless, many conversations we examined mentioned how badly the existing appeal process functioned, echoing what happened in online communities~\cite{myers2018censored}. When appealing decisions where their account were restricted on a GAI online tool due to content moderation, users often faced long wait times for a final decision, or sometimes, never received a response at all. Users mentioned that this situation could occur when appealing to the moderation system failures as well. A user who was previously banned by OpenAI because of moderation from DALL-E wrote: \textit{``It took a solid 5 months to get my account back. It’s a massive problem that your account can be terminated for innocent mistakes, and it takes that long to recover it.''} 
Meanwhile, users who appealed or submitted feedback regarding moderated content also reported that the process was sometimes ineffective, not only in having their creation requests reconsidered but also in obtaining an explanation for the moderation decision. A Midjourney user wrote: \textit{``Now I try to appeal and it just gives me this: `Sorry! Our AI moderator thinks this prompt is probably against our community standards. Please review our current community standards...'''} Because of the broken appeal pipeline, some users would not even consider appealing even when facing severe problems. As a Midjourney user who received massive arbitrary false positive moderation on their creations acknowledged: \textit{``I know I can contact support, but that takes time and effort every time.''} 

\section{Discussion}
Our policy analysis shows content moderation policies in GAI online tools are fairly comprehensive but lack details (\cref{sec:findings1}). Through investigating user perceptions, we also reveal that content moderation in GAI online tools succeeds in blocking the majority of malicious content generation but users feel that they fail in making justified moderation decisions and providing post-moderation support (\cref{sec:findings2}). 
Based on our findings, we outline implications for improving GAI content moderation policies and enforcement below.

\subsection{Improve Moderation Policy Outline}
%Our findings in policy analysis revealed similarities and differences in structure and composition between content moderation policies in GAI online tools and those in online communities (Table \ref{tab:policycomparision}). We identified similar pitfalls in policies of both types of platforms, while also noticing the uniqueness of policymaking in GAI online tools that need attention in policymaking. 
Similar to content moderation policies in online communities, we found content moderation policies in GAI online tools were scattered on multiple pages (\cref{susec: analysis1}) and that most tools lack a dedicated section outlining all rules for prohibited inputs and outputs (\cref{subsec: criteriapresent}). 
%We found an amount of content moderation criteria is delivered through AUP, as the community guidelines in online communities do~\cite{10.1145/3406865.3418312}. However, AUP themselves also often lack a standardized location, structure, or even consistent naming across different products~\cite{klyman2024acceptable}, unlike community guidelines. 
We also found that content moderation policies for GAI online tools and their related online communities (e.g., image generation tools that are built with or integrated into online communities) are mixed up together (\cref{subsec: criteriapresent}), making it difficult to distinguish policies for the GAI from policies on posting to a corresponding online community. Given the trend of integrating GAI services into online communities (e.g.,\cite{socialmediatoday_tiktok_gen_ai}), this could deepen the problem for both external regulators to check legal compliance at scale and users to seek guidance on content generation using a GAI product, similar to the case of online communities with unstructured policies~\cite{schaffner2024community}. We, therefore, recommend that GAI products \textbf{establish a unified and clear policy structure} for presenting content moderation rules. This should include establishing dedicated policy pages for all GAI content moderation policies, and separating moderation rules for GAI input and output from those on GAI-associated online communities. %This should include dedicated policy or support pages that comprehensively outline or link all content moderation-related information for input and output using GAI tools. %Additionally, it is crucial to distinguish content moderation policies specific to GAI from those applicable to other activities such as interacting in a GAI-related online community.

Moreover, we found content moderation policies in GAI online tools commonly lack detailed user-driven moderation methods or provide information on user appeal when they are moderated (\cref{subsec:usermethod}, \cref{subsec: userappeal}). Online communities heavily rely on collective user action for detecting problematic content to overcome challenges of platform-driven moderation, such as inaccuracies and biases in algorithmic detection, the sheer volume of user-generated content that burdens human moderators, and socially contextual boundaries of issues like misinformation and discrimination~\cite{schaffner2024community, seering2020reconsidering, 10.1145/3476059}. In GAI products, similar issues exist in moderation systems, with the challenges further amplified by the unpredictability~\cite{10.1145/3548606.3560599, gehman2020realtoxicityprompts} and biases~\cite{gillespie2024generative} of GAI outputs. %Research has shown that LLM outputs containing inherent biases can not be prohibited by moderation mechanisms alone but also by human prompting~\cite{fan2024user}. 
Researchers have been advocating for collective user involvement in GAI safety, e.g., through user-driven GAI output auditing~\cite{deng2025weaudit} and user-driven value alignment~\cite{fan2024user}. Therefore, we strongly recommend GAI products \textbf{set up detailed procedures with clear steps for user feedback on problematic GAI output} that are not detected by moderation systems. \textbf{A robust user appeal pipeline} can also provide a feedback mechanism to collect information on questionable moderation decisions, beyond supporting users after moderation.

\subsection{Balance Usability and Safety in Moderation}

We found users spoke of how moderation systems successfully blocked malicious AIGC creative tasks in GAI online tools in public discussions (\cref{subsec:success}). Simultaneously, we perceived that user-experienced failures of moderation systems in AIGC creative tasks were often linked to the censorship of normal requests (\cref{subsec:systemfail}). In online communities, the trade-offs between free speech and maintaining community safety as well as the boundaries for content moderation have been long debated (e.g., \cite{langvardt2017regulating}). For instance, artists argued that censorship of nudity in artwork prevents them from contributing to the creative community and engaging with other artists~\cite{riccio2024exposed}. Our findings highlighted a similar tension between GAI tool usability, creativity, and safety in content moderation enforcement. 

One direct approach to promote user safety while ensuring GAI usability is to \textbf{improve the accuracy and context awareness of GAI moderation systems}, as stakeholders have constantly worked on (e.g., \cite{zhang2024controllable}). We found user frustrations about outright user request denials and GAI output blocks that completely disrupt normal use, especially when caused by moderation system decision-making failures (\cref{subsec:systemfail}). Therefore, GAI products could \textbf{employ soft moderation for user input and GAI output}. In online communities, soft moderation is when a platform issues a warning of content or decreases the content visibility in recommendation feeds, instead of direct content removals or account bans~\cite{singhal2023sok, zannettou2021won}. Similar strategies have already been adopted in real-world GAI products, with evidence indicating that they can improve both usability and user satisfaction. A prior work found that users are most frustrated with direct denials without explanations when making LLM requests, but are more satisfied when a diverted task from the original request is fulfilled~\cite{wester2024ai}. We, furthermore, recommend a primary use of soft moderation if there is a chance of moderation system failures. That is, when there is a high likelihood of falsely moderating input or output, the system should mask the content, issue a warning, or generate a modified output, instead of completely blocking or removing the content.

Our findings also highlight user-reported issues of moderation systems making decisions without sufficient awareness of task context (\cref{subsec:systemfail}). In response, GAI products could also \textbf{deploy personalized content moderation guardrails}, where users can adjust contextualized input and output filters for different tasks. This could be similar to the approach used in online communities for users to configure their safety preferences and personalized filters to customize if, and to what extent, they are exposed to disturbing content~\cite{jhaver2023personalizing}. We note that personalized content moderation guardrails carry potential risks, including abuse by malicious users and misuse by minors. Therefore, deploying this feature requires careful consideration of the extent to which guardrails can be personalized, limiting access to users with a positive usage history (i.e., no malicious use or jailbreaking attempts), and incorporating features like parental controls.

\subsection{Make Moderation Pipeline Transparent}
When using GAI online tools for AIGC creative tasks, users not only encounter moderation system failures that result in unreasonable decisions but also experience frustration due to the lack of explanations for these decisions, as well as the absence of an effective user appeal process (\cref{subsec:providerfail}). The lack of transparency and user support in the moderation pipeline has already been widely discussed in the context of online communities~\cite{myers2018censored, vaccaro2020end}. We, however, consider this situation to be more detrimental to the user experience of GAI products, given how moderation systems in GAI products function at every stage from user input through output generation progress, to output endpoint. This is unlike online communities that solely host and moderate user-generated content.

%As previously mentioned, GAI involves the interplay of user input, generated output, and the output generation process, unlike online communities that solely host and moderate user-generated content. Therefore, moderation mechanisms in GAI online tools are implemented at every stage of output generation: from the input starting point (external check for input filtering), through input processing and output generation (training safer models to produce less problematic outputs), to the output endpoint (external check for output filtering), as disclosed in policies (\cref{subsec:toolmethod}).

Therefore, we suggest that GAI products implement a supportive pipeline for users after they are moderated. They should \textbf{provide clear explanations and customer support after moderation} about which stage their generation request was moderated and which moderation criteria were possibly breached. This would help users to interpret at which stage they were moderated and reason about whether a moderation decision results from randomly generated malicious output or stems from inherent biases and inaccuracies within the moderation algorithm. This will also help users understand why their content or accounts were restricted, offering transparency and clarity in the decision-making process.

Additionally, we found user complaints about the insufficient details of moderation methods in policies, which aligns with our findings from the policy analysis (\cref{subsec:toolmethod}, \cref{subsec:providerfail}). To address this issue, GAI product service providers should \textbf{further elaborate on how the moderation system operates in each stage}. This is especially critical for automatic methods, as content moderation criteria in policies and metrics used by algorithmic moderation are sometimes misaligned~\cite{arora2023detecting}. If an approach using implicit rules, like blocking input with specific words, is adopted, policies should also present these implicit rules, such as making the complete ban word list visible to users.

\vspace{-0.7em}
\section{Conclusion}
We analyzed GAI online tool content moderation policies, finding that these policies resemble those of online communities but place emphasis on governing inputs and outputs, employing unique detection methods and response strategies for problematic content. While policies in GAI online tools comprehensively outline content moderation practices from criteria to enforcement, they lack provisions for user-driven moderation methods and appeal pipelines. We also analyzed public discussions about the GAI online tool moderation for AIGC creative tasks on Reddit. We found that while moderation systems effectively block malicious AIGC creation, users frequently discuss instances where these systems fail to make justified decisions and do not provide enough information about moderation decisions. We suggest that the GAI product policy structure can be improved, with more information on users' roles in content moderation and user appeals, and providing better explanations for why and when moderation occurs. Future work can study additional GAI products and conduct user studies to build on our findings.
\section*{Ethical Considerations}
We identified no ethical concerns in the policy analysis study, as all the web pages we collected were publicly available, not linked to any individuals, and gathered without the use of scraping tools. We also did not require ethical reviews for the policy analysis study. The Reddit study was reviewed and approved by our institutional IRB before data collection. To protect the privacy of users who made Reddit posts and comments we collected, we removed all identifiable information and slightly adjusted the wording of each quote we reported in the paper. 

\section*{Open Science}
We have made the following outcomes from our paper publicly available: the policy dataset (screenshots of pages collected for Study 1: Policy Analysis) with the analysis outcome (the codebook and annotated policy segments); and the Reddit dataset (Reddit posts collected for Study 2: Reddit Study) with the codebook for analysis. This artifact can be found at \url{https://doi.org/10.6084/m9.figshare.29257187}. 

We have not made comments corresponding to the collected Reddit posts publicly available due to the ethical considerations stated above. 

\bibliographystyle{plain}
\bibliography{main}

\begin{thebibliography}{10}

\bibitem{ahn2023splintering}
Soyun Ahn, Jeeyun Baik, and Clara~Sol Krause.
\newblock Splintering and centralizing platform governance: how facebook adapted its content moderation practices to the political and legal contexts in the united states, germany, and south korea.
\newblock {\em Information, Communication \& Society}, 26(14):2843--2862, 2023.

\bibitem{appel2024generative}
Ruth~Elisabeth Appel.
\newblock Generative ai regulation can learn from social media regulation.
\newblock {\em arXiv preprint arXiv:2412.11335}, 2024.

\bibitem{armstrong1997place}
David Armstrong, Ann Gosling, John Weinman, and Theresa Marteau.
\newblock The place of inter-rater reliability in qualitative research: An empirical study.
\newblock {\em Sociology}, 31(3):597--606, 1997.

\bibitem{arora2023detecting}
Arnav Arora, Preslav Nakov, Momchil Hardalov, Sheikh~Muhammad Sarwar, Vibha Nayak, Yoan Dinkov, Dimitrina Zlatkova, Kyle Dent, Ameya Bhatawdekar, Guillaume Bouchard, et~al.
\newblock Detecting harmful content on online platforms: what platforms need vs. where research efforts go.
\newblock {\em ACM Computing Surveys}, 56(3):1--17, 2023.

\bibitem{bai2022traininghelpfulharmlessassistant}
Yuntao Bai, Andy Jones, Kamal Ndousse, Amanda Askell, Anna Chen, Nova DasSarma, Dawn Drain, Stanislav Fort, Deep Ganguli, Tom Henighan, et~al.
\newblock Training a helpful and harmless assistant with reinforcement learning from human feedback.
\newblock {\em arXiv preprint arXiv:2204.05862}, 2022.

\bibitem{binns2017like}
Reuben Binns, Michael Veale, Max Van~Kleek, and Nigel Shadbolt.
\newblock Like trainer, like bot? inheritance of bias in algorithmic content moderation.
\newblock In {\em Social Informatics: 9th International Conference, SocInfo 2017, Oxford, UK, September 13-15, 2017, Proceedings, Part II 9}, pages 405--415. Springer, 2017.

\bibitem{buckley2022censorship}
Nicole Buckley and Joseph~S Schafer.
\newblock 'censorship-free'platforms: Evaluating content moderation policies and practices of alternative social media.
\newblock 2022.

\bibitem{cao2023comprehensive}
Yihan Cao, Siyu Li, Yixin Liu, Zhiling Yan, Yutong Dai, Philip~S Yu, and Lichao Sun.
\newblock A comprehensive survey of ai-generated content (aigc): A history of generative ai from gan to chatgpt.
\newblock {\em arXiv preprint arXiv:2303.04226}, 2023.

\bibitem{10.1145/3134666}
Eshwar Chandrasekharan, Umashanthi Pavalanathan, Anirudh Srinivasan, Adam Glynn, Jacob Eisenstein, and Eric Gilbert.
\newblock You can't stay here: The efficacy of reddit's 2015 ban examined through hate speech.
\newblock {\em Proc. ACM Hum.-Comput. Interact.}, 1(CSCW), December 2017.

\bibitem{chen2023pathway}
Chen Chen, Jie Fu, and Lingjuan Lyu.
\newblock A pathway towards responsible ai generated content.
\newblock {\em arXiv preprint arXiv:2303.01325}, 2023.

\bibitem{cnbc_chatgpt_image_generations}
CNBC.
\newblock Chatgpt blocked 250,000 image generations of presidential candidates.
\newblock \url{https://www.cnbc.com/2024/11/08/chatgpt-blocked-250000-image-generations-of-presidential-candidates.html}, 2024.
\newblock Accessed: 2025-01-08.

\bibitem{dai2024safesora}
Josef Dai, Tianle Chen, Xuyao Wang, Ziran Yang, Taiye Chen, Jiaming Ji, and Yaodong Yang.
\newblock Safesora: Towards safety alignment of text2video generation via a human preference dataset.
\newblock {\em arXiv preprint arXiv:2406.14477}, 2024.

\bibitem{dai2023saferlhfsafereinforcement}
Josef Dai, Xuehai Pan, Ruiyang Sun, Jiaming Ji, Xinbo Xu, Mickel Liu, Yizhou Wang, and Yaodong Yang.
\newblock Safe rlhf: Safe reinforcement learning from human feedback.
\newblock {\em arXiv preprint arXiv:2310.12773}, 2023.

\bibitem{deng2025weaudit}
Wesley~Hanwen Deng, Claire Wang, Howard~Ziyu Han, Jason~I Hong, Kenneth Holstein, and Motahhare Eslami.
\newblock Weaudit: Scaffolding user auditors and ai practitioners in auditing generative ai.
\newblock {\em arXiv preprint arXiv:2501.01397}, 2025.

\bibitem{eslami2016first}
Motahhare Eslami, Karrie Karahalios, Christian Sandvig, Kristen Vaccaro, Aimee Rickman, Kevin Hamilton, and Alex Kirlik.
\newblock First i" like" it, then i hide it: Folk theories of social feeds.
\newblock In {\em Proceedings of the 2016 cHI conference on human factors in computing systems}, pages 2371--2382, 2016.

\bibitem{fan2024user}
Xianzhe Fan, Qing Xiao, Xuhui Zhou, Jiaxin Pei, Maarten Sap, Zhicong Lu, and Hong Shen.
\newblock User-driven value alignment: Understanding users' perceptions and strategies for addressing biased and discriminatory statements in ai companions.
\newblock {\em arXiv preprint arXiv:2409.00862}, 2024.

\bibitem{10.1145/2675133.2675234}
Casey Fiesler, Jessica~L. Feuston, and Amy~S. Bruckman.
\newblock Understanding copyright law in online creative communities.
\newblock In {\em Proceedings of the 18th ACM Conference on Computer Supported Cooperative Work \& Social Computing}, CSCW '15, page 116–129, New York, NY, USA, 2015. Association for Computing Machinery.

\bibitem{fiesler2018reddit}
Casey Fiesler, Jialun Jiang, Joshua McCann, Kyle Frye, and Jed Brubaker.
\newblock Reddit rules! characterizing an ecosystem of governance.
\newblock In {\em Proceedings of the International AAAI Conference on Web and Social Media}, volume~12, 2018.

\bibitem{10.1145/2818048.2819931}
Casey Fiesler, Cliff Lampe, and Amy~S. Bruckman.
\newblock Reality and perception of copyright terms of service for online content creation.
\newblock In {\em Proceedings of the 19th ACM Conference on Computer-Supported Cooperative Work \& Social Computing}, CSCW '16, page 1450–1461, New York, NY, USA, 2016. Association for Computing Machinery.

\bibitem{fiesler2024remember}
Casey Fiesler, Michael Zimmer, Nicholas Proferes, Sarah Gilbert, and Naiyan Jones.
\newblock Remember the human: A systematic review of ethical considerations in reddit research.
\newblock {\em Proceedings of the ACM on Human-Computer Interaction}, 8(GROUP):1--33, 2024.

\bibitem{flexos_generative_ai_top150}
FlexOS.
\newblock [report] generative ai top 150: The world's most used ai tools (feb 2024).
\newblock \url{https://www.flexos.work/learn/generative-ai-top-150}, 2025.
\newblock Accessed: 2025-01-08.

\bibitem{forbes_chatgpt_ai_tool_sector}
Forbes.
\newblock New research shows chatgpt reigns supreme in ai tool sector.
\newblock \url{https://www.forbes.com/sites/chriswestfall/2023/11/16/new-research-shows-chatgpt-reigns-supreme-in-ai-tool-sector/}, 2023.
\newblock Accessed: 2025-01-08.

\bibitem{garcia2024generative}
Francisco~Jos{\'e} Garc{\'\i}a-Pe{\~n}alvo.
\newblock Generative artificial intelligence and education: An analysis from multiple perspectives.
\newblock {\em Education in the Knowledge Society}, 25:e31942, 2024.

\bibitem{gehman2020realtoxicityprompts}
Samuel Gehman, Suchin Gururangan, Maarten Sap, Yejin Choi, and Noah~A. Smith.
\newblock {R}eal{T}oxicity{P}rompts: Evaluating neural toxic degeneration in language models.
\newblock In Trevor Cohn, Yulan He, and Yang Liu, editors, {\em Findings of the Association for Computational Linguistics: EMNLP 2020}, pages 3356--3369, Online, November 2020. Association for Computational Linguistics.

\bibitem{gillespie2017governance}
Tarleton Gillespie.
\newblock Governance of and by platforms.
\newblock {\em SAGE handbook of social media}, pages 254--278, 2017.

\bibitem{gillespie2018custodians}
Tarleton Gillespie.
\newblock {\em Custodians of the Internet: Platforms, Content Moder ation, and the Hidden Decisions that Shape Social Media}.
\newblock Yale University Press, 2018.

\bibitem{gillespie2024generative}
Tarleton Gillespie.
\newblock Generative ai and the politics of visibility.
\newblock {\em Big Data \& Society}, 11(2):20539517241252131, 2024.

\bibitem{gillespie2020expanding}
Tarleton Gillespie, Patricia Aufderheide, Elinor Carmi, Ysabel Gerrard, Robert Gorwa, Ariadna Matamoros-Fern{\'a}ndez, Sarah~T Roberts, Aram Sinnreich, and Sarah Myers~West.
\newblock Expanding the debate about content moderation: Scholarly research agendas for the coming policy debates.
\newblock {\em Internet Policy Review}, 9(4):1--29, 2020.

\bibitem{github_awesome_generative_ai}
Github.
\newblock Awesome generative ai.
\newblock \url{https://github.com/steven2358/awesome-generative-ai}, 2025.
\newblock Accessed: 2025-01-08.

\bibitem{goldman2021content}
Eric Goldman.
\newblock Content moderation remedies.
\newblock {\em Mich. Tech. L. Rev.}, 28:1, 2021.

\bibitem{gomez2024algorithmic}
Juan~Felipe Gomez, Caio Machado, Lucas~Monteiro Paes, and Flavio Calmon.
\newblock Algorithmic arbitrariness in content moderation.
\newblock In {\em The 2024 ACM Conference on Fairness, Accountability, and Transparency}, pages 2234--2253, 2024.

\bibitem{hacker2023regulating}
Philipp Hacker, Andreas Engel, and Marco Mauer.
\newblock Regulating chatgpt and other large generative ai models.
\newblock In {\em Proceedings of the 2023 ACM Conference on Fairness, Accountability, and Transparency}, pages 1112--1123, 2023.

\bibitem{haimson2021disproportionate}
Oliver~L Haimson, Daniel Delmonaco, Peipei Nie, and Andrea Wegner.
\newblock Disproportionate removals and differing content moderation experiences for conservative, transgender, and black social media users: Marginalization and moderation gray areas.
\newblock {\em Proceedings of the ACM on Human-Computer Interaction}, 5(CSCW2):1--35, 2021.

\bibitem{han2022aligning}
Shengnan Han, Eugene Kelly, Shahrokh Nikou, and Eric-Oluf Svee.
\newblock Aligning artificial intelligence with human values: reflections from a phenomenological perspective.
\newblock {\em AI \& SOCIETY}, pages 1--13, 2022.

\bibitem{hua2024generative}
Yiqing Hua, Shuo Niu, Jie Cai, Lydia~B Chilton, Hendrik Heuer, and Donghee~Yvette Wohn.
\newblock Generative ai in user-generated content.
\newblock In {\em Extended Abstracts of the CHI Conference on Human Factors in Computing Systems}, pages 1--7, 2024.

\bibitem{inan2023llama}
Hakan Inan, Kartikeya Upasani, Jianfeng Chi, Rashi Rungta, Krithika Iyer, Yuning Mao, Michael Tontchev, Qing Hu, Brian Fuller, Davide Testuggine, et~al.
\newblock Llama guard: Llm-based input-output safeguard for human-ai conversations.
\newblock {\em arXiv preprint arXiv:2312.06674}, 2023.

\bibitem{10.1145/3359294}
Shagun Jhaver, Darren~Scott Appling, Eric Gilbert, and Amy Bruckman.
\newblock "did you suspect the post would be removed?": Understanding user reactions to content removals on reddit.
\newblock {\em Proc. ACM Hum.-Comput. Interact.}, 3(CSCW), November 2019.

\bibitem{10.1145/3359252}
Shagun Jhaver, Amy Bruckman, and Eric Gilbert.
\newblock Does transparency in moderation really matter? user behavior after content removal explanations on reddit.
\newblock {\em Proc. ACM Hum.-Comput. Interact.}, 3(CSCW), November 2019.

\bibitem{jhaver2023personalizing}
Shagun Jhaver, Alice~Qian Zhang, Quan~Ze Chen, Nikhila Natarajan, Ruotong Wang, and Amy~X Zhang.
\newblock Personalizing content moderation on social media: User perspectives on moderation choices, interface design, and labor.
\newblock {\em Proceedings of the ACM on Human-Computer Interaction}, 7(CSCW2):1--33, 2023.

\bibitem{ji2024beavertails}
Jiaming Ji, Mickel Liu, Josef Dai, Xuehai Pan, Chi Zhang, Ce~Bian, Boyuan Chen, Ruiyang Sun, Yizhou Wang, and Yaodong Yang.
\newblock Beavertails: Towards improved safety alignment of llm via a human-preference dataset.
\newblock {\em Advances in Neural Information Processing Systems}, 36, 2024.

\bibitem{ji2023ai}
Jiaming Ji, Tianyi Qiu, Boyuan Chen, Borong Zhang, Hantao Lou, Kaile Wang, Yawen Duan, Zhonghao He, Jiayi Zhou, Zhaowei Zhang, et~al.
\newblock Ai alignment: A comprehensive survey.
\newblock {\em arXiv preprint arXiv:2310.19852}, 2023.

\bibitem{10.1145/3406865.3418312}
Jialun~'Aaron' Jiang, Skyler Middler, Jed~R. Brubaker, and Casey Fiesler.
\newblock Characterizing community guidelines on social media platforms.
\newblock In {\em Companion Publication of the 2020 Conference on Computer Supported Cooperative Work and Social Computing}, CSCW '20 Companion, page 287–291, New York, NY, USA, 2020. Association for Computing Machinery.

\bibitem{jin2024jailbreaking}
Haibo Jin, Andy Zhou, Joe~D Menke, and Haohan Wang.
\newblock Jailbreaking large language models against moderation guardrails via cipher characters.
\newblock {\em arXiv preprint arXiv:2405.20413}, 2024.

\bibitem{juneja2020through}
Prerna Juneja, Deepika Rama~Subramanian, and Tanushree Mitra.
\newblock Through the looking glass: Study of transparency in reddit's moderation practices.
\newblock {\em Proceedings of the ACM on Human-Computer Interaction}, 4(GROUP):1--35, 2020.

\bibitem{kiesler2012regulating}
Sara Kiesler, Robert Kraut, Paul Resnick, and Aniket Kittur.
\newblock Regulating behavior in online communities.
\newblock {\em Building successful online communities: Evidence-based social design}, 1:4--2, 2012.

\bibitem{klonick2017new}
Kate Klonick.
\newblock The new governors: The people, rules, and processes governing online speech.
\newblock {\em Harv. L. Rev.}, 131:1598, 2017.

\bibitem{klyman2024acceptable}
Kevin Klyman.
\newblock Acceptable use policies for foundation models.
\newblock In {\em Proceedings of the AAAI/ACM Conference on AI, Ethics, and Society}, volume~7, pages 752--767, 2024.

\bibitem{kou2024community}
Yubo Kou, Renkai Ma, Zinan Zhang, Yingfan Zhou, and Xinning Gui.
\newblock Community begins where moderation ends: Peer support and its implications for community-based rehabilitation.
\newblock In {\em Proceedings of the CHI Conference on Human Factors in Computing Systems}, pages 1--18, 2024.

\bibitem{langvardt2017regulating}
Kyle Langvardt.
\newblock Regulating online content moderation.
\newblock {\em Geo. LJ}, 106:1353, 2017.

\bibitem{liu2024safetydpo}
Runtao Liu, Chen~I Chieh, Jindong Gu, Jipeng Zhang, Renjie Pi, Qifeng Chen, Philip Torr, Ashkan Khakzar, and Fabio Pizzati.
\newblock Safetydpo: Scalable safety alignment for text-to-image generation.
\newblock {\em arXiv preprint arXiv:2412.10493}, 2024.

\bibitem{lyons2022s}
Henrietta Lyons, Senuri Wijenayake, Tim Miller, and Eduardo Velloso.
\newblock What’s the appeal? perceptions of review processes for algorithmic decisions.
\newblock In {\em Proceedings of the 2022 CHI Conference on Human Factors in Computing Systems}, pages 1--15, 2022.

\bibitem{ma2021advertiser}
Renkai Ma and Yubo Kou.
\newblock " how advertiser-friendly is my video?": Youtuber's socioeconomic interactions with algorithmic content moderation.
\newblock {\em Proceedings of the ACM on Human-Computer Interaction}, 5(CSCW2):1--25, 2021.

\bibitem{ma2023users}
Renkai Ma, Yue You, Xinning Gui, and Yubo Kou.
\newblock How do users experience moderation?: A systematic literature review.
\newblock {\em Proceedings of the ACM on Human-Computer Interaction}, 7(CSCW2):1--30, 2023.

\bibitem{mahomed2024auditing}
Yaaseen Mahomed, Charlie~M Crawford, Sanjana Gautam, Sorelle~A Friedler, and Dana{\"e} Metaxa.
\newblock Auditing gpt's content moderation guardrails: Can chatgpt write your favorite tv show?
\newblock In {\em The 2024 ACM Conference on Fairness, Accountability, and Transparency}, pages 660--686, 2024.

\bibitem{markov2023holistic}
Todor Markov, Chong Zhang, Sandhini Agarwal, Florentine~Eloundou Nekoul, Theodore Lee, Steven Adler, Angela Jiang, and Lilian Weng.
\newblock A holistic approach to undesired content detection in the real world.
\newblock In {\em Proceedings of the AAAI Conference on Artificial Intelligence}, volume~37, pages 15009--15018, 2023.

\bibitem{10.1145/3512965}
J.~Nathan Matias, Austin Hounsel, and Nick Feamster.
\newblock Software-supported audits of decision-making systems: Testing google and facebook's political advertising policies.
\newblock {\em Proc. ACM Hum.-Comput. Interact.}, 6(CSCW1), April 2022.

\bibitem{10.1145/3359174}
Nora McDonald, Sarita Schoenebeck, and Andrea Forte.
\newblock Reliability and inter-rater reliability in qualitative research: Norms and guidelines for cscw and hci practice.
\newblock {\em Proc. ACM Hum.-Comput. Interact.}, 3(CSCW), November 2019.

\bibitem{moran2022folk}
Rachel~E Moran, Izzi Grasso, and Kolina Koltai.
\newblock Folk theories of avoiding content moderation: How vaccine-opposed influencers amplify vaccine opposition on instagram.
\newblock {\em Social Media+ Society}, 8(4):20563051221144252, 2022.

\bibitem{myers2018censored}
Sarah Myers~West.
\newblock Censored, suspended, shadowbanned: User interpretations of content moderation on social media platforms.
\newblock {\em New Media \& Society}, 20(11):4366--4383, 2018.

\bibitem{ouyang2022traininglanguagemodelsfollow}
Long Ouyang, Jeffrey Wu, Xu~Jiang, Diogo Almeida, Carroll Wainwright, Pamela Mishkin, Chong Zhang, Sandhini Agarwal, Katarina Slama, Alex Ray, et~al.
\newblock Training language models to follow instructions with human feedback.
\newblock {\em Advances in neural information processing systems}, 35:27730--27744, 2022.

\bibitem{rando2022red}
Javier Rando, Daniel Paleka, David Lindner, Lennart Heim, and Florian Tram{\`e}r.
\newblock Red-teaming the stable diffusion safety filter.
\newblock {\em arXiv preprint arXiv:2210.04610}, 2022.

\bibitem{riccio2024exploring}
Piera Riccio, Georgina Curto, and Nuria Oliver.
\newblock Exploring the boundaries of content moderation in text-to-image generation.
\newblock {\em arXiv preprint arXiv:2409.17155}, 2024.

\bibitem{riccio2024exposed}
Piera Riccio, Thomas Hofmann, and Nuria Oliver.
\newblock Exposed or erased: Algorithmic censorship of nudity in art.
\newblock In {\em Proceedings of the CHI Conference on Human Factors in Computing Systems}, pages 1--17, 2024.

\bibitem{roberts2019behind}
Sarah~T Roberts.
\newblock {\em Behind the screen: Content Moderation in the Shadows of Social Media}.
\newblock Yale University Press, 2019.

\bibitem{samuelson2023generative}
Pamela Samuelson.
\newblock Generative ai meets copyright.
\newblock {\em Science}, 381(6654):158--161, 2023.

\bibitem{writerbuddy_ai_industry_analysis}
Sujan Sarkar.
\newblock Ai industry analysis: 50 most visited ai tools and their 24b+ traffic behavior.
\newblock \url{https://writerbuddy.ai/blog/ai-industry-analysis/}, 2023.
\newblock Accessed: 2025-01-08.

\bibitem{saunders2018saturation}
Benjamin Saunders, Julius Sim, Tom Kingstone, Shula Baker, Jackie Waterfield, Bernadette Bartlam, Heather Burroughs, and Clare Jinks.
\newblock Saturation in qualitative research: exploring its conceptualization and operationalization.
\newblock {\em Quality \& quantity}, 52:1893--1907, 2018.

\bibitem{schaffner2024community}
Brennan Schaffner, Arjun~Nitin Bhagoji, Siyuan Cheng, Jacqueline Mei, Jay~L Shen, Grace Wang, Marshini Chetty, Nick Feamster, Genevieve Lakier, and Chenhao Tan.
\newblock " community guidelines make this the best party on the internet": An in-depth study of online platforms' content moderation policies.
\newblock In {\em Proceedings of the CHI Conference on Human Factors in Computing Systems}, pages 1--16, 2024.

\bibitem{schmitt2024implications}
Vera Schmitt, Jakob Tesch, Eva Lopez, Tim Polzehl, Aljoscha Burchardt, Konstanze Neumann, Salar Mohtaj, and Sebastian M{\"o}ller.
\newblock Implications of regulations on large generative ai models in the super-election year and the impact on disinformation.
\newblock In {\em Proceedings of the Workshop on Legal and Ethical Issues in Human Language Technologies@ LREC-COLING 2024}, pages 28--38, 2024.

\bibitem{schramowski2023safe}
Patrick Schramowski, Manuel Brack, Bj{\"o}rn Deiseroth, and Kristian Kersting.
\newblock Safe latent diffusion: Mitigating inappropriate degeneration in diffusion models.
\newblock In {\em Proceedings of the IEEE/CVF Conference on Computer Vision and Pattern Recognition}, pages 22522--22531, 2023.

\bibitem{schulman2017proximalpolicyoptimizationalgorithms}
John Schulman, Filip Wolski, Prafulla Dhariwal, Alec Radford, and Oleg Klimov.
\newblock Proximal policy optimization algorithms.
\newblock {\em arXiv preprint arXiv:1707.06347}, 2017.

\bibitem{seering2020reconsidering}
Joseph Seering.
\newblock Reconsidering self-moderation: the role of research in supporting community-based models for online content moderation.
\newblock {\em Proceedings of the ACM on Human-Computer Interaction}, 4(CSCW2):1--28, 2020.

\bibitem{10.1145/3548606.3560599}
Wai~Man Si, Michael Backes, Jeremy Blackburn, Emiliano De~Cristofaro, Gianluca Stringhini, Savvas Zannettou, and Yang Zhang.
\newblock Why so toxic? measuring and triggering toxic behavior in open-domain chatbots.
\newblock In {\em Proceedings of the 2022 ACM SIGSAC Conference on Computer and Communications Security}, CCS '22, page 2659–2673, New York, NY, USA, 2022. Association for Computing Machinery.

\bibitem{singhal2023sok}
Mohit Singhal, Chen Ling, Pujan Paudel, Poojitha Thota, Nihal Kumarswamy, Gianluca Stringhini, and Shirin Nilizadeh.
\newblock Sok: Content moderation in social media, from guidelines to enforcement, and research to practice.
\newblock In {\em 2023 IEEE 8th European Symposium on Security and Privacy (EuroS\&P)}, pages 868--895. IEEE, 2023.

\bibitem{statista_generative_ai_usage}
Statista.
\newblock Intention of generative artificial intelligence (gai) usage by adults in the united states as of august 2023, by type.
\newblock \url{https://www.statista.com/statistics/1461998/usa-generative-ai-usage-intention-by-type/}, 2024.
\newblock Accessed: 2025-01-08.

\bibitem{statista_generative_ai_us}
Statista.
\newblock Use of generative artificial intelligence (ai) programs in the united states in 2023, by use case.
\newblock \url{https://www.statista.com/statistics/1413836/use-of-generative-ai-us/}, 2024.
\newblock Accessed: 2025-01-08.

\bibitem{suzor2019lawless}
Nicolas~P Suzor.
\newblock {\em Lawless: The secret rules that govern our digital lives}.
\newblock Cambridge University Press, 2019.

\bibitem{299651}
Madiha Tabassum, Alana Mackey, Ashley Schuett, and Ada Lerner.
\newblock Investigating moderation challenges to combating hate and harassment: The case of {Mod-Admin} power dynamics and feature misuse on reddit.
\newblock In {\em 33rd USENIX Security Symposium (USENIX Security 24)}, pages 37--54, Philadelphia, PA, August 2024. USENIX Association.

\bibitem{socialmediatoday_tiktok_gen_ai}
Social~Media Today.
\newblock Tiktok adds more generative ai features.
\newblock \url{https://www.socialmediatoday.com/news/tiktok-adds-gen-ai-image-tools-caption-suggestions/736361/}, 2025.
\newblock Accessed: 2025-01-19.

\bibitem{tourkochoriti2023digital}
Ioanna Tourkochoriti.
\newblock The digital services act and the eu as the global regulator of the internet.
\newblock {\em Chi. J. Int'l L.}, 24:129, 2023.

\bibitem{touvron2023llama2openfoundation}
Hugo Touvron, Louis Martin, Kevin Stone, Peter Albert, Amjad Almahairi, Yasmine Babaei, Nikolay Bashlykov, Soumya Batra, Prajjwal Bhargava, Shruti Bhosale, et~al.
\newblock Llama 2: Open foundation and fine-tuned chat models.
\newblock {\em arXiv preprint arXiv:2307.09288}, 2023.

\bibitem{uscode_title15_section9401}
{U.S. House of Representatives}.
\newblock United states code: Title 15, section 9401 (preliminary edition).
\newblock \url{https://uscode.house.gov/view.xhtml?req=(title:15%20section:9401%20edition:prelim)}, 2025.
\newblock Accessed: 2025-01-08.

\bibitem{vaccaro2020end}
Kristen Vaccaro, Christian Sandvig, and Karrie Karahalios.
\newblock " at the end of the day facebook does what itwants" how users experience contesting algorithmic content moderation.
\newblock {\em Proceedings of the ACM on human-computer interaction}, 4(CSCW2):1--22, 2020.

\bibitem{10.1145/3476059}
Kristen Vaccaro, Ziang Xiao, Kevin Hamilton, and Karrie Karahalios.
\newblock Contestability for content moderation.
\newblock {\em Proc. ACM Hum.-Comput. Interact.}, 5(CSCW2), October 2021.

\bibitem{wang2024moderator}
Peiran Wang, Qiyu Li, Longxuan Yu, Ziyao Wang, Ang Li, and Haojian Jin.
\newblock Moderator: Moderating text-to-image diffusion models through fine-grained context-based policies.
\newblock In {\em Proceedings of the 2024 on ACM SIGSAC Conference on Computer and Communications Security}, pages 1181--1195, 2024.

\bibitem{wei2022finetunedlanguagemodelszeroshot}
Jason Wei, Maarten Bosma, Vincent~Y Zhao, Kelvin Guu, Adams~Wei Yu, Brian Lester, Nan Du, Andrew~M Dai, and Quoc~V Le.
\newblock Finetuned language models are zero-shot learners.
\newblock {\em arXiv preprint arXiv:2109.01652}, 2021.

\bibitem{wei2024understanding}
Yiluo Wei and Gareth Tyson.
\newblock Understanding the impact of ai-generated content on social media: The pixiv case.
\newblock In {\em Proceedings of the 32nd ACM International Conference on Multimedia}, pages 6813--6822, 2024.

\bibitem{wester2024ai}
Joel Wester, Tim Schrills, Henning Pohl, and Niels van Berkel.
\newblock “as an ai language model, i cannot”: Investigating llm denials of user requests.
\newblock In {\em Proceedings of the CHI Conference on Human Factors in Computing Systems}, pages 1--14, 2024.

\bibitem{wu2023ai}
Jiayang Wu, Wensheng Gan, Zefeng Chen, Shicheng Wan, and Hong Lin.
\newblock Ai-generated content (aigc): A survey.
\newblock {\em arXiv preprint arXiv:2304.06632}, 2023.

\bibitem{wu2023finegrainedhumanfeedbackgives}
Zeqiu Wu, Yushi Hu, Weijia Shi, Nouha Dziri, Alane Suhr, Prithviraj Ammanabrolu, Noah~A Smith, Mari Ostendorf, and Hannaneh Hajishirzi.
\newblock Fine-grained human feedback gives better rewards for language model training.
\newblock {\em Advances in Neural Information Processing Systems}, 36:59008--59033, 2023.

\bibitem{298254}
Zhiyuan Yu, Xiaogeng Liu, Shunning Liang, Zach Cameron, Chaowei Xiao, and Ning Zhang.
\newblock Don{\textquoteright}t listen to me: Understanding and exploring jailbreak prompts of large language models.
\newblock In {\em 33rd USENIX Security Symposium (USENIX Security 24)}, pages 4675--4692, Philadelphia, PA, August 2024. USENIX Association.

\bibitem{zannettou2021won}
Savvas Zannettou.
\newblock " i won the election!": an empirical analysis of soft moderation interventions on twitter.
\newblock In {\em Proceedings of the international AAAI conference on web and social media}, volume~15, pages 865--876, 2021.

\bibitem{zhang2024controllable}
Jingyu Zhang, Ahmed Elgohary, Ahmed Magooda, Daniel Khashabi, and Benjamin Van~Durme.
\newblock Controllable safety alignment: Inference-time adaptation to diverse safety requirements.
\newblock {\em arXiv preprint arXiv:2410.08968}, 2024.

\bibitem{zhang2023siren}
Yue Zhang, Yafu Li, Leyang Cui, Deng Cai, Lemao Liu, Tingchen Fu, Xinting Huang, Enbo Zhao, Yu~Zhang, Yulong Chen, et~al.
\newblock Siren's song in the ai ocean: a survey on hallucination in large language models.
\newblock {\em arXiv preprint arXiv:2309.01219}, 2023.

\bibitem{zhou2023synthetic}
Jiawei Zhou, Yixuan Zhang, Qianni Luo, Andrea~G Parker, and Munmun De~Choudhury.
\newblock Synthetic lies: Understanding ai-generated misinformation and evaluating algorithmic and human solutions.
\newblock In {\em Proceedings of the 2023 CHI Conference on Human Factors in Computing Systems}, pages 1--20, 2023.

\end{thebibliography}

\appendix

\section*{Appendix}
See the page below for appendix tables.

\begin{table*}[]
\renewcommand\arraystretch{1.1}
\centering
\resizebox{1.9\columnwidth}{!}{
\begin{tabular}{p{0.2\textwidth}|p{0.2\textwidth}p{0.2\textwidth}|p{0.2\textwidth}p{0.2\textwidth}}
\hline
                 & \textbf{Dataset}& & \textbf{Sample for Analysis} & \\
\hline
Subreddit                 & Post        & Comment       & Post              & Comment             \\
\hline
r/chatgpt        & 335         & 17090         & 38                & 2069                \\
r/midjourney     & 216         & 4383          & 24                & 473                 \\
r/dalle2         & 240         & 5423          & 33                & 686                 \\
r/claudeAI       & 169         & 4230          & 17                & 280                 \\
r/Bard           & 142         & 2223          & 16                & 322                 \\
r/dalle          & 14          & 75            & 1                 & 6                   \\
r/perplexity\_ai & 7           & 41            & 1                 & 3                   \\
\hline
Sum              & 1123        & 33465         & 130               & 3839\\
\hline
\end{tabular}}
\captionsetup{justification=centering}
\caption{Statistics of Reddit dataset and random sample for analysis}\label{tab:reddit}
\end{table*}

\begin{table*}[]
    \centering
    \renewcommand\arraystretch{1.2}
    \resizebox{1.9\columnwidth}{!}{
    \begin{tabular}{p{0.17\textwidth}p{0.8\textwidth}}
    \toprule
    \textbf{Restrictions}     &\textbf{Examples of Over-sensitive Input Moderation} \\
    \midrule
    Harmful Content     &\textit{I tried earlier to generate an image of a mother and child but it refused to do so. I took the child out of the request and it worked great. Not being able to generate images of minors in compromising situations is one thing, but to filter it out entirely is too restricting.} (Gemini, r/Bard) \\
    Content that Violates Other’s Rights     &\textit{I'm just trying to brainstorm some lyric ideas for AI music and half the time it works fine other half Claude tells me they can't write music. I'm not even telling it to copy someone's style or anything. I usually give it a rough few lines I wrote and then it spits nothing out.} (Claude, r/ClaudeAI)\\
    Sexual     &\textit{I had issues simply with ``the 2 characters' foreheads are touching in a display of tenderness and affection.''} (ChatGPT's DALL-E, r/chatgpt)\\
    Violence     & \textit{I tried to make a kitchen image. Cutting board was banned due to cutting.} (Midjourney, r/midjourney) \\
    \bottomrule
    \end{tabular}}
    \caption{Content moderation criteria and corresponding examples of false-positive input moderation. Note that the criteria `content that is not appropriate for everyone' is split into `sexual' and `violence' when elaborating examples.}
    \label{tab:overreaction} %\mc{showing frequency? need to show what is new here and that we learned from Reddit - haven't thought}
\end{table*}

\begin{table*}[]
    \centering
    \renewcommand\arraystretch{1.2}
    \resizebox{1.9\columnwidth}{!}{
    \begin{tabular}{p{0.17\textwidth}p{0.8\textwidth}}
    \toprule
    \textbf{Restrictions}     &\textbf{Examples of Over-restrictions} \\
    \midrule
    Harmful Content     &\textit{Then you mention the bank robber likes to kick puppies. Or that they're actively targeting a bank in a spot with a vulnerable population, which is a very logical thing for a bank robber to do [...]  ``Oh no, I can't write anything condoning kicking puppies or assaulting vulnerable people.''} (Claude, r/claudeAI) \\
    Content that Violates Other’s Rights     &\textit{I tried to put a photo of myself in the editor and it said `This violates our terms of service.’} (Midjourney, r/midjourney) \\
    Misleading Content     &\textit{I asked it to generate an image of the 6th Army sieging Kingslanding and it refused because the 6th Army is real and Kingslanding is fantasy.} (Gemini, r/Bard) \\
    Sexual     &\textit{All of my generated articles were just removed after I included a passage mentioning an adult work on sexuality and gender studies in my prompt.} (ChatGPT, r/chatgpt) \\
    Violence     &\textit{The censorship has gotten so out of hand that I can't generate pictures with a halloween theme because they have `blood' in the prompt.} (Midjourney, r/midjourney) \\
    Value Alignment     &\textit{I have a fight scene and if the hero ever even remotely has anything bad happen, they instantly rebound with great courage and become a role model to everyone, vanquishing the baddie for all time.} (ChatGPT, r/chatgpt) \\
    \bottomrule
    \end{tabular}}
    \caption{Content moderation criteria and corresponding examples of over-restrictions. Note that the criteria `content that is not appropriate for everyone' is split into `sexual' and `violence' when elaborating examples. Value alignment is also included here.}
    \label{tab:overrestriction}
\end{table*}

%\theendnotes

\end{document}